\documentclass[epj,nopacs,fleqn]{svjour}
\pdfoutput=1
\usepackage{undertilde,graphicx,url,amsmath,amssymb,slashed,cite}
\usepackage[usenames,dvipsnames]{color}

\begin{document}

\newcommand{\nc}{\newcommand}
\nc{\postscript}[2]{\setlength{\epsfxsize}{#2\hsize}\centerline{\epsfbox{#1}}}
\nc{\be}{\begin{equation}}   \nc{\ee}{\end{equation}}
\nc{\beq}{\begin{eqnarray}}   \nc{\eeq}{\end{eqnarray}}
\nc{\bea}{\begin{align}}   \nc{\eea}{\end{align}}
\nc{\bit}{\begin{itemize}}    \nc{\eit}{\end{itemize}}
\nc{\ben}{\begin{enumerate}}  \nc{\een}{\end{enumerate}}
\nc{\bce}{\begin{center}}     \nc{\ece}{\end{center}}
\nc{\nn}{\nonumber}
\nc\bpm{\begin{pmatrix}}      \nc\epm{\end{pmatrix}} 
\def\bsp#1\esp{\begin{split}#1\end{split}}

\newcommand{\etc}{{\it etc.}}
\newcommand{\ie}{{\it i.e.}}
\newcommand{\eg}{{\it e.g.}}

\newcommand\sss{\scriptscriptstyle}
\newcommand\cw{\cos\theta_{\sss W}}
\newcommand\cwsq{\hat c^2_{\sss W}}
\newcommand\bcw{\bar c_{\sss W}}
\newcommand\bcb{\bar c_{\sss B}}
\newcommand\bchw{\bar c_{\sss HW}}
\newcommand\bchb{\bar c_{\sss HB}}
\newcommand\bcbb{\bar c_{\sss BB}}
\newcommand\hcw{\hat c_{\sss W}}
\newcommand\hsw{\hat s_{\sss W}}
\newcommand\swsq{\hat s^2_{\sss W}}
\newcommand\swcu{\hat s^3_{\sss W}}
\newcommand\swqu{\hat s^4_{\sss W}}
\newcommand\cwsm{\cos\tilde\theta_{\sss W}}
\newcommand\swsqsm{\sin^2\tilde\theta_{\sss W}}
\newcommand\cwsqsm{\cos^2\tilde\theta_{\sss W}}
\newcommand\cwdsm{\cos2\tilde\theta_{\sss W}}
\newcommand\sw{\sin\theta_{\sss W}}
\newcommand\mw{m_{\sss W}}
\newcommand\mz{m_{\sss Z}}
\newcommand\mh{m_{\sss H}}
 \def\lra#1{\overset{\text{\scriptsize$\leftrightarrow$}}{#1}}

\newcommand{\red}[1]{\color{red} #1 \color{black}}


\title{Electroweak Higgs boson production in the standard model effective field theory beyond leading order in QCD}

\author{
C\'eline Degrande\inst{1,}\thanks{celine.degrande@cern.ch},
Benjamin Fuks\inst{2,3,}\thanks{fuks@lpthe.jussieu.fr},
Kentarou Mawatari\inst{4,5,}\thanks{kentarou.mawatari@lpsc.in2p3.fr},
Ken Mimasu\inst{6,7,}\thanks{ken.mimasu@uclouvain.be},
Ver\'onica Sanz\inst{6,}\thanks{v.sanz@sussex.ac.uk}
}

\institute{
 CERN, Theory Division, Geneva 23 CH-1211, Switzerland
\and
UPMC Univ.~Paris 06,  Sorbonne Universit\'es, UMR 7589, LPTHE, 75005 Paris, France
\and
 CNRS, UMR 7589, LPTHE, 75005 Paris, France
\and
 Laboratoire de Physique Subatomique et de Cosmologie, Universit\'e Grenoble-Alpes, CNRS/IN2P3,\\ 53 Avenue des Martyrs, 38026 Grenoble, France 
\and
 Theoretische Natuurkunde and IIHE/ELEM, Vrije Universiteit Brussel,
 and International Solvay Institutes,\\
 Pleinlaan 2, 1050 Brussels, Belgium
\and
 Department of Physics and Astronomy, University of Sussex, 
 Brighton BN1 9QH, UK
 \and
 Centre for Cosmology, Particle Physics and Phenomenology (CP3), Universit\'e 
 catholique de Louvain, \\
 Chemin du Cyclotron, 1348 Louvain-la-Neuve, Belgium
}

\date{} 

\abstract{
We study the impact of dimension-six operators of the standard
model effective field theory relevant for vector-boson fusion and associated Higgs
boson production at the LHC. We present predictions at the next-to-leading order
accuracy in QCD that include matching to parton showers and that rely on fully
automated simulations. We show the importance of the subsequent reduction of the
theoretical uncertainties in improving the possible discrimination between
effective field theory and standard model results, and we demonstrate
that the range of the Wilson coefficient values allowed by a global fit to LEP
and LHC Run~I data can be further constrained by LHC Run~II future results.
}

\titlerunning{
Electroweak Higgs boson production in the Standard Model EFT beyond leading order in QCD
}   
\authorrunning{C.~Degrande et al.}

\maketitle

\vspace*{-14cm}
\noindent 
IPPP/16/80, LPSC16198
\vspace*{12cm}


\section{Introduction}

The LHC Run~I and early Run~II data have not yet put forward any strong evidence
of physics beyond the standard model (SM) and limits on new states have instead
been pushed to higher and higher energies. As a consequence, the effective field
theory (EFT) extension of the SM (SMEFT) has become increasingly relevant. The
SMEFT is built from the SM symmetries and degrees of freedom (including the
Higgs sector) by adding new operators of dimension higher than four to the
SM Lagrangian. Being a tool to parameterise the search for new anomalous
interactions, it is fully complementary to direct searches for new particles.
Interpreting data in the context of the SMEFT hence allows us to be sensitive to
new physics beyond the current energy reach of the LHC in a model-independent
way.

The formulation of the
effective Lagrangian restricted to operators of dimension of at most six
relies on the definition of a complete and non-redundant operator
basis~\cite{Grzadkowski:2010es,Contino:2013kra,Gupta:2014rxa} and should
additionally include the
translations among the possible choices~\cite{Falkowski:2015wza}. This has been
intensively discussed and will be soon reported by the Higgs cross section
working group~\cite{YR4}. 
Moreover,
since we try to observe small deviations from the SM, precise theoretical
predictions are required both in the SM and in the SMEFT framework.
The accumulation of LHC data
and the subsequent precision obtained indeed call for a similar accuracy on the
theoretical side, which demands the inclusion of higher-order corrections.

What we present in this paper is a part of the current theoretical activities aiming for
precision predictions for electroweak Higgs-boson production at the LHC, 
{\it i.e.} Higgs boson production in association with a weak boson (VH) and via vector-boson fusion (VBF).
In this context,
NLO+PS (next-to-leading order plus parton-shower) matched predictions for VH and VBF production in the SM
have been released both in the MC@NLO~\cite{Frixione:2002ik,LatundeDada:2009rr,%
Frixione:2013mta} and {\sc Pow\-heg}~\cite{Nason:2004rx,Hamilton:2009za,%
Nason:2009ai} frameworks, and merged
NLO samples describing VH production including up to one additional jet have
been generated in both the {\sc Powheg}~\cite{Luisoni:2013kna} and
{\sc Sherpa}~\cite{Goncalves:2015mfa} platforms. NLO QCD corrections along with
the inclusion of anomalous interactions have been further investigated for
VH~\cite{Campanario:2014lza} and VBF~\cite{Hankele:2006ma} Higgs boson
production, and matched to parton showers in the Higgs
characterisation
framework~\cite{Artoisenet:2013puc,Maltoni:2013sma}. Finally, electroweak
corrections as well as anomalous coupling effects for VH production have been included in the {\sc Hawk} program~\cite{Denner:2011id} that also contains 
NLO QCD contributions.
In contrast, fixed-order predictions are known to a higher accuracy for
both VH and VBF SM Higgs production processes~\cite{Kumar:2014uwa,%
Campbell:2016jau,Dreyer:2016oyx,Cacciari:2015jma}.

In the SMEFT framework (in contrast to the anomalous coupling
approach), the VH process has been studied
at the NLO+PS accuracy within the {\sc Powheg-Box}
framework~\cite{Mimasu:2015nqa}, where a subset of the 59 independent dimension-six operators was taken into account.
In this paper, similarly, we consider five operators which are relevant for VH production.
Firstly, we independently provide NLO+PS predictions for the VH process
by using a different framework via a joint use of
{\sc FeynRules}~\cite{Alloul:2013bka}, {\sc Nlo\-CT}~\cite{Degrande:2014vpa}
and {\sc Mad\-Graph5\_a\-MC@NLO} ({\sc MG5\_a\-MC}) \cite{Alwall:2014hca}
programs. This approach
provides a fully automatic procedure linking the model Lagrangian to event
generation matched to parton showers at NLO. 
Our work hence not only independently
validates the previous results obtained with {\sc Powheg-Box} but also includes the additional benefits stemming
from the flexibility of the {\sc FeynRules} program. As a result, one can exploit generators
like {\sc MG5\_aMC} for simulating any desired process at the NLO+PS accuracy (i.e. the VBF process in our case) for which the same operators play a role.
This is the second part of our paper, which presents the first SMEFT results for this process.
Although we only present results
for a couple of benchmark scenarios motivated by global fit results, our predictions can be straightforwardly
generalised to any scenario by using our public Universal {\sc FeynRules} Output
(UFO) model~\cite{FR-HEL:Online} within
{\sc MG5\_aMC}~\cite{Christensen:2009jx,Degrande:2011ua,deAquino:2011ub}.

We emphasise that, following some recent results in the $t\bar{t}H$ channel~\cite{Maltoni:2016yxb}, 
this work represents a step towards a complete SMEFT operator basis
implementation for Higgs physics at the NLO QCD accuracy,
which will be beneficial to both the theoretical and experimental communities.  

In Sect.~\ref{sec:theory} we provide the
necessary theoretical ingredients to calculate NLO-QCD corrections for VH and VBF Higgs production in the SMEFT.
We also discuss current constraints on the Wilson coefficients originating
from a LEP and LHC Run~I global fit analysis with which we inform our benchmark
point selection. In Sect.~\ref{sec:sim} we describe our setup for NLO
computations matched to parton showers. We present our numerical results
in Sects.~\ref{sec:WH} and~\ref{sec:VBF}, and also assess the validity of the EFT
given the current constraints on the Wilson coefficients. We asses the reach of the future
LHC reach in Sect.~\ref{sec:reach}, before concluding in
Sect.~\ref{sec:conclusion}. Practical information for event simulation and
model validations are provided in the appendix.

\section{Theoretical framework\label{sec:theory}}
\subsection{Model description}
In the SM of particle physics, the elementary particles and their
interactions are described by a quantum field theory based on the
$SU(3)_C\times SU(2)_L\times U(1)_Y$ gauge symmetry. The vector fields mediating
the gauge interactions lie in the adjoint representation of the relevant gauge group,
\begin{align}
  SU(3)_C &\to G= ({\utilde{\bf 8}},{\utilde{\bf 1}},0) \ ,\nn\\
  SU(2)_L &\to W= ({\utilde{\bf 1}},{\utilde{\bf 3}},0) \ ,\nn\\
  U(1)_Y  &\to B= ({\utilde{\bf 1}},{\utilde{\bf 1}},0) \ ,
\end{align}
where the notations for the representation refer to the full SM
symmetry group. The chiral content of the theory is defined by three generations
of left-handed and right-handed quark ($Q_L$, $u_R$ and $d_R$) and lepton ($L_L$
and $e_R$) fields whose representation under the SM gauge group is
given by
\begin{align}
  Q_L &=\bpm u_L\\d_L \epm = \big(\utilde{\bf 3},\utilde{\bf 2},\frac16\big)\ ,
  \nn\\
  u_R &= \big(\utilde{\bf 3},\utilde{\bf 1}, \frac23\big)\ ,  \ \
  d_R = \big(\utilde{\bf 3},\utilde{\bf 1},-\frac13\big)\ ,\nn\\
  L_L &=\bpm\nu_L\\\ell_L\epm=\big(\utilde{\bf 1},\utilde{\bf 2},-\frac12\big)\ ,
  \ \
  e_R = \big(\utilde{\bf 1},\utilde{\bf 1},-1\big)\ .
\end{align}
The Higgs sector contains a single $SU(2)_L$ doublet of fields
that is responsible for the breaking of electroweak symmetry,
\begin{align}
  \Phi = \bpm -i G^+ \\ \frac{1}{\sqrt{2}} \Big[ v + h + i G^0\Big] \epm = 
      \big(\utilde{\bf 1}, \utilde{\bf 2},\frac12\big) \ ,
\end{align}
where the components of the $\Phi$ doublet are given in terms of the
physical Higgs field $h$ shifted by its vacuum expectation value $v$ and the
Goldstone bosons $G^\pm$ and $G^0$ that are eaten by the weak bosons to give them their
longitudinal degree of freedom.

In the EFT framework, new physics is expected to appear at a
scale $\Lambda$ large enough so that the new degrees of freedom can be
integrated out. As a result, the SM Lagrangian $\mathcal{L}_{\rm SM}$
is supplemented by higher-dimensional operators $\mathcal{O}_i$ parameterising all
effects beyond the SM,
\be
  {\cal L} = {\cal L}_{\rm SM} +
    \sum_{n=1}^\infty \sum_i \frac{\bar c_{ni}}{\Lambda^n} \mathcal{O}_{ni}.
\ee
Restricting ourselves to operators of dimension six,
the most general gauge-invariant Lagrangian $\mathcal{L}$
has been known for a long time \cite{Burges:1983zg,Leung:1984ni,Buchmuller:1985jz}
and can be expressed in a suitable form by choosing a convenient basis of
independent operators $\mathcal{O}_i$
~\cite{Grzadkowski:2010es,Contino:2013kra,Gupta:2014rxa}.
In this work, we focus on five specific, bosonic operators,\footnote{The relevant fermionic operators are also considered in, e.g., \cite{Ge:2016zro}.} which are relevant to the VH
and VBF processes, taken from the strongly
interacting light Higgs (SILH) basis
~\cite{Giudice:2007fh,Contino:2013kra,Alloul:2013naa},%
\footnote{Although the $W$-boson mass $\mw$ and $v$ are usually used as
expansion parameters in this basis, our
model explicitly uses a cutoff scale $\Lambda$. For all our numerical results,
we set \mbox{$\Lambda=\mw$}. We also point out a relative factor 2 difference in our definition of ${\cal O}_{\sss W}$ and ${\cal O}_{\sss HW}$ with respect to
Refs.~\cite{Giudice:2007fh,Contino:2013kra,Alloul:2013naa}.}
\begin{align}
  \mathcal{L} = & \ \mathcal{L}_{\rm SM} +
    \frac{g'^2}{4\Lambda^2}\bar c_{\sss BB}
       \Phi^\dagger \Phi B_{\mu\nu}B^{\mu\nu} \nn\\
  & \ + \frac{ig}{2\Lambda^2}\bar c_{\sss W}
       \big[\Phi^\dag T_{2k} \overleftrightarrow{D}_\mu \Phi \big]
       D_\nu  W^{k,\mu \nu} \nn\\
  & \ + \frac{ig'}{2\Lambda^2}\bar c_{\sss B}
      \big[\Phi^\dag \overleftrightarrow{D}_\mu \Phi \big]
	        \partial_\nu  B^{\mu \nu} \nn\\
  & \ + \frac{ig}{\Lambda^2}\bar c_{\sss HW} \big[D_\mu \Phi^\dag T_{2k}
      D_\nu \Phi\big] W^{k,\mu \nu} \nn\\
  & \ + \frac{ig'}{\Lambda^2}\bar c_{\sss HB} \big[D_\mu \Phi^\dag D_\nu \Phi
      \big] B^{\mu \nu} \ .
\label{eqn:lhelc}
\end{align}
The Wilson coefficients $\bar c$
are free parameters, \mbox{$T_{2k}$}  are the generators of $SU(2)$ (with \mbox{${\rm Tr}(T_{2k}T_{2l})=\delta_{kl}/2$}) in the fundamental representation
and the Hermitian derivative operators are defined by
\begin{align}
  \Phi^\dag {\overleftrightarrow D}_\mu \Phi =&\
     \Phi^\dag (D_\mu\Phi)-(D_\mu\Phi^\dag)\Phi\ ,\nn\\
  \Phi^\dag T_{2k} {\overleftrightarrow D}_\mu \Phi = &\
     \Phi^\dag T_{2k}(D_\mu\Phi)-(D_\mu\Phi^\dag) T_{2k}\Phi \ .
\end{align}
In our conventions, the gauge-covariant derivatives and the gauge field
strength tensors read
\begin{align}
  W^k_{\mu\nu} &= \partial_\mu W^k_\nu - \partial_\nu W^k_\mu + g \epsilon_{ij}{}^k \ W^i_\mu W^j_\nu\ ,\nn\\
  B_{\mu\nu} &= \partial_\mu B_\nu - \partial_\nu B_\mu\ , \nn\\
  D_\rho W^k_{\mu\nu} &= \partial_\rho W^k_{\mu\nu} + g \epsilon_{ij}{}^k W^i_\rho W_{\mu\nu}^j \ ,\nn\\
  D_\mu\Phi &= \partial_\mu \Phi -  i g T_{2k} W_\mu^k \Phi - \frac12 i g' B_\mu \Phi  \ , 
\end{align}
where 
$\epsilon_{ij}{}^k$ are the structure constants of $SU(2)$. In addition, $g$
and $g'$ denote the coupling constants of $SU(2)_L$ and $U(1)_Y$ respectively.

After the breaking of the electroweak symmetry down to electromagnetism, the
weak and hypercharge gauge ei\-gen\-states mix to the physical $W$-boson, $Z$-boson
and the photon $A$,
\begin{align}
  W_\mu^\pm &= {1\over \sqrt{2}}  (W_\mu^1 \mp i W^2_\mu)\ , \nn\\
  \bpm Z_\mu\\ A_\mu \epm &=
   \bpm \hcw & -\hsw \\ \hsw & \phantom{-}\hcw\epm \bpm W_\mu^3 \\ B_\mu\epm \ .
\label{eqn:weakshifts} 
\end{align}
We have introduced in this expression the sine and cosine of the weak
mixing angle \mbox{$\hsw\equiv\sin\hat\theta_{\sss W}$} and
\mbox{$\hcw\equiv\cos\hat\theta_{\sss W}$}
which diagonalise the neutral electroweak gauge boson mass matrix.
The higher-dimensional operators of Eq.~\eqref{eqn:lhelc} induce a modification
of the gauge boson kinetic terms that become, in the mass basis and
after integration by parts,
\begin{align}
  &{\cal L}_{\rm kin} =
    -\frac12 \bigg[1 -\frac{g ^2 v^2 \bar c_{\sss W}}{4 \Lambda^2} \bigg] W^{+}_{\mu\nu} W^{-\mu\nu}\nn\\
  &\ -
    \frac14 \bigg[1-\frac{g^{\prime2} v^2 \bar c_{\sss BB}}{2\cwsq \Lambda^2}\bigg]
    A_{\mu\nu} A^{\mu\nu} \nn\\
  &\ -
  \frac14 \bigg[1 -\frac{g^2 v^2 \bar c_{\sss W}}{4 \Lambda^2}-\frac{g^{\prime2} v^2 \bar c_{\sss B}}{2 \Lambda^2} -
       \frac{g^{\prime2} \swsq v^2 \bar c_{\sss BB}}{2 \Lambda^2}\bigg] Z_{\mu\nu}  Z^{\mu\nu}\nn\\
  &\ + 
   \frac{v^2}{\Lambda^2}\bigg[ \frac{g^2 \hsw \bar c_{\sss W}}{16 \hcw }  - \frac{g^{\prime 2} \hcw \bar c_{\sss B}}{8 \hsw }- \frac{g^{\prime2} \hsw\hcw \bar c_{\sss BB}}{4} \bigg] A_{\mu\nu} Z^{\mu\nu} \ ,
\end{align}
where $W^{\pm}_{\mu\nu}$, $Z_{\mu\nu}$ and $A_{\mu\nu}$ denote the $W$-boson,
$Z$-boson and photon field strength tensors, respectively. 
Con\-se\-quen\-tly, canonical normalisation has to be restored by redefining
the electroweak boson fields,
\begin{align}
   W_\mu &\to
      \bigg[1 + \frac{g^2 v^2 \bar c_{\sss W}}{8 \Lambda^2}\bigg]W_\mu\ ,
     \nn\\
   Z_\mu &\to  \bigg[1
        + \frac{g^2 v^2 \bar c_{\sss W}}{8 \Lambda^2} 
        + \frac{g^{\prime2} v^2 \bar c_{\sss B}}{4 \Lambda^2}
        + \frac{g^{\prime2} \swsq v^2 \bar c_{\sss BB}}{4 \Lambda^2} \bigg]Z_\mu
        \nn \\
     &\quad +  \frac{v^2}{\Lambda^2}\bigg[
          \frac{g^2 \hsw \hcw \bar c_{\sss W}}{8} 
        - \frac{g^{\prime2} \hsw \hcw \bar c_{\sss BB}}{4}
    \bigg]A_\mu\ , \nn\\
   A_\mu &\to  \nn
      \bigg[1 + \frac{g^{\prime2} \cwsq v^2 \bar c_{\sss BB}}{4 \Lambda^2}
      \bigg]A_\mu \\
      &\quad  +  \frac{v^2}{\Lambda^2}\bigg[
          \frac{g^2 \swcu \bar c_{\sss W}}{8 \hcw}
        - \frac{g^{\prime 2} \hcw \bar c_{\sss B}}{4 \hsw}
        - \frac{g^{\prime2} \hsw\hcw \bar c_{\sss BB}}{4}\bigg]Z_\mu\ .
\label{eqn:fieldredef}
\end{align}
We have made use here of the freedom related to the
removal of the photon and $Z$-boson mixing terms induced by the higher-order
operators. This mixing can indeed be absorbed either in a photon field
redefinition, or in a $Z$-boson field redefinition, or in both (as in
Eq.~\eqref{eqn:fieldredef}).  In order to minimise the modification of the weak
interactions with respect to the SM, we additionally redefine the
weak and hypercharge coupling constants
\be
  g \to \frac{e}{\hsw}\bigg[ 1 - \frac{e^2 v^2 \bar c_{\sss W}}{8 \hsw^2\Lambda^2}\bigg]\ , \ \
  g' \to  \frac{e}{\hcw}\bigg[ 1 - \frac{e^2 v^2 \bar c_{\sss BB}}{4\hcw^2 \Lambda^2}\bigg] \ .
\ee
As a result of this choice, the
relations between the measured values for the electroweak inputs and all
internal electroweak parameters are simplified.
The $Z$-boson mass $\mz$ is now given by
\be
  \mz = \frac{e v}{2 \hsw\hcw} \bigg[ 1 + \frac{e^2 v^2}{8 \cwsq \Lambda^2}
        \Big(\cwsq \bar c_{\sss W} + 2 \bar c_{\sss B}\Big) \bigg] \ ,
\label{eqn:weakmasses}\ee
while the photon stays massless and the expression of the  $W$-boson mass $\mw$ is unchanged respect to the SM one.

We define the electroweak sector of the theory in terms of the Fermi coupling
constant $G_\mathrm{F}$ as extracted from the muon decay data, the measured $Z$-boson
mass $\mz$ and the electromagnetic coupling constant $\alpha$ in the low-energy
limit of the Compton scattering. The vacuum expectation value of the Higgs field
can therefore be derived from the Fermi constant as in the SM,
$ v^2 = 1/(\sqrt{2} G_\mathrm{F})\,$.
After the field redefinitions of Eq.~\eqref{eqn:fieldredef}, the electromagnetic
interactions of the fermions to the photon field turn out to be solely modified by
the ${\cal O}_{\sss W}$ operator, so that the electromagnetic
coupling constant $e$ is related to the input parameter $\alpha$ as
\be
  e = \sqrt{4 \pi \alpha} \bigg[1 +  \frac{\pi \alpha v^2 \bar c_{\sss W}}{2 \Lambda^2} \bigg]\ .
 \label{eqn:e_alpha}
\ee
Furthermore, the shift in the cosine of the Weinberg mixing angle $\cos\hat\theta_W$
can be derived, at first order in $1/\Lambda^2$, from the $Z$-boson mass
relation of Eq.~\eqref{eqn:weakmasses} along with Eq.~\eqref{eqn:e_alpha},
\be
  \hcw^2 = \tilde c_{\sss W}^2 - \frac{2\pi\alpha \tilde s_{\sss W}^2 v^2 }{\tilde c_{\sss 2W} \Lambda^2}
    \big[ \tilde c_{\sss W}^2 \bar c_{\sss W} +
    \bar c_{\sss B}\big]\ ,
\label{eqn:cosw}
\ee
with
\mbox{$\tilde c_{\sss 2W}\equiv\cos{2\tilde\theta_{\sss W}}$},
\mbox{$\tilde s_{\sss W}\equiv\sin\tilde\theta_{\sss W}$} and
\be
  \tilde c_{\sss W}^2 \equiv \cwsqsm
    = \frac12\bigg[ 1 + \sqrt{1-\frac{4\pi\alpha v^2}{\mz^2}}\bigg] \ .
\ee

As a consequence, the $\bar c_{\sss W}$ and $\bar c_{\sss B}$ parameters are
constrained by the measurement of the $W$-boson mass and by the $Z$-boson
decay data. Those constraints can nonetheless be modified if other dimension-six
operators are added to the Lagrangian of Eq.~\eqref{eqn:lhelc}.

\setlength{\tabcolsep}{10pt}
\renewcommand{\arraystretch}{1.6}
\begin{table*}
  \center
  \begin{tabular}{c|lll}
  \hline
    Eq.~\eqref{eq:L_hvv} & Our conventions & Ref.~\cite{Alloul:2013naa} (HEL) & Ref.~\cite{Artoisenet:2013puc} (HC)\\
  \hline
  $g_{\sss h\gamma\gamma}$ &
    $a_{\sss H} - \frac{e^2 v}{\Lambda^2} \bar c_{\sss BB} $ &
    $a_{\sss H} - \frac{8 g \hsw^2}{\mw}\bar c_{\sss \gamma} $ &
    $c_\alpha \kappa_{\sss H\gamma\gamma} g_{\sss H\gamma\gamma}$\\
  $g^{(1)}_{\sss hzz}$ &
  $\frac{e^2v}{2\hsw^2\hcw^2\Lambda^2}\left[
      \hcw^2\bchw+2\hsw^2\bchb-2\hsw^4\bcbb
      \right]$ 
  &
    $\frac{2 g}{\cwsq \mw} \Big[\cwsq \bar c_{\sss HW}+  \swsq \bar c_{\sss HB} - 4 \swqu \bar c_{\sss \gamma} \Big]$ &
    $\frac{1}{\Lambda} c_\alpha \kappa_{\sss HZZ}$ \\
  $g^{(2)}_{\sss hzz}$ &
    $\frac{e^2 v}{4 \swsq\cwsq \Lambda^2} \Big[ \cwsq (\bar c_{\sss HW} + \bar c_{\sss W})+ 2\swsq (\bar c_{\sss HB}+\bar c_{\sss B} )\Big]$   &
    $\frac{g}{\cwsq \mw} \Big[\cwsq (\bar c_{\sss HW} +\bar c_{\sss W})  + \swsq(\bar c_{\sss HB} + \bar c_{\sss B} ) \Big]$&
    $\frac{1}{\Lambda} c_\alpha \kappa_{\sss H\partial Z}$ \\
  $g^{(3)}_{\sss hzz}$  &
    $\frac{g^2 v}{2 \cwsq} + \frac{e^4 v^3}{8 \swsq\hcw^4 \Lambda^2} \Big[\cwsq \bar c_{\sss W} + 2 \bar c_{\sss B}\Big]$ &
    $\frac{g \mw}{\cwsq} \Big[ 1 +  \frac{8\swqu}{\cwsq} \bar c_{\sss\gamma} \Big]$ &
    $c_\alpha \kappa_{\rm SM} g_{\sss HZZ}$\\
  $g^{(1)}_{\sss hz\gamma}$ & $a'_{\sss H}$+
    $\frac{e^2v}{4 \hsw\hcw \Lambda^2} \Big[ \bar c_{\sss HW} - 2\bar c_{\sss HB} + 4 \swsq \bar c_{\sss BB} \Big]$       &
    $\frac{g \hsw}{\hcw \mw} \Big[  \bar c_{\sss HW} - \bar c_{\sss HB} + 8  \swsq \bar c_{\sss \gamma} \Big]$ &
    $c_\alpha \kappa_{\sss HZ\gamma} g_{\sss HZ\gamma}$ \\
  $g^{(2)}_{\sss hz\gamma}$ &
    $\frac{e^2v}{4 \hsw \hcw \Lambda^2} \Big[ \bar c_{\sss HW} +\bar c_{\sss W} -2 (\bar c_{\sss BB}+ \bar c_{\sss B}
    )\Big]$       &
    $\frac{g \hsw}{\hcw \mw} \Big[  \bar c_{\sss HW}+ \bar c_{\sss W} - (\bar c_{\sss HB}  + \bar c_{\sss B})\Big]$ &
    $\frac{1}{\Lambda} c_\alpha \kappa_{\sss H\partial \gamma}$ \\
  $g^{(1)}_{\sss hww}$ &
    $\frac{e^2 v}{2 \swsq \Lambda^2} \bar c_{\sss HW}$ &
    $\frac{2 g}{\mw} \bar c_{\sss HW}$ &
    $\frac{1}{\Lambda} c_\alpha \kappa_{\sss HWW}$ \\
  $g^{(2)}_{\sss hww}$ &
    $\frac{e^2 v}{4 \swsq \Lambda^2} \Big[ \bar c_{\sss HW} +\bar c_{\sss W}  \Big]$ &
    $\frac{g}{\mw} \Big[ \bar c_{\sss HW} + \bar c_{\sss W} \Big]$ &
    $\frac{1}{\Lambda} c_\alpha \kappa_{\sss H\partial W}$ \\
  $g^{(3)}_{\sss hww}$ &
    $\frac{g^2 v}{2}  $&
    $g \mw$ &
    $c_\alpha \kappa_{\rm SM} g_{\sss HWW}$ \\ 
	\hline
\end{tabular}
  \caption{New physics effects in three-point interactions involving a Higgs boson and electroweak gauge bosons.
    The loop-induced SM contributions to the Higgs-boson couplings
    to two photons $a_{\sss H}$ and to one $Z$-boson and one photon $a'_{\sss H}$ 
    have been explicitly indicated.
	}
  \label{tab:L3conv}
\end{table*}
\renewcommand{\arraystretch}{1}

In unitary gauge and rotating all field to the mass basis, all three-point
interactions involving a single (physical) Higgs boson and a pair of electroweak gauge bosons are given by
\begin{align}
  &{\cal L}_{\rm hvv} =
    - \frac{1}{4} g_{\sss h\gamma\gamma} A_{\mu\nu} A^{\mu\nu} h 
     - \frac{1}{4} g_{\sss hzz}^{(1)} Z_{\mu\nu} Z^{\mu\nu} h \nn\\
    &\quad - g_{\sss hzz}^{(2)} Z_\nu \partial_\mu Z^{\mu\nu} h
    + \frac{1}{2} g_{\sss hzz}^{(3)} Z_\mu Z^\mu h \nn\\
    &\quad - \frac{1}{2} g_{\sss hz\gamma}^{(1)} Z_{\mu\nu} A^{\mu\nu} h
    - g_{\sss hz\gamma}^{(2)} Z_\nu \partial_\mu A^{\mu\nu} h \nn\\
    &\quad - \frac{1}{2} g_{\sss hww}^{(1)} W_{\mu\nu}^+ W^{-\mu\nu} h 
   - g_{\sss hww}^{(2)} \, \Big[ W_\nu^+ \partial_\mu W^{-\mu\nu} h + {\rm h.c.} \Big] \nn\\
    &\quad +  g_{\sss hww}^{(3)} W_\mu^+  W^{-\mu} h \ ,
\label{eq:L_hvv}
\end{align}
where integration by parts
has been used to reduce the number of independent Lorentz structures.
Table~\ref{tab:L3conv} shows the relation between the couplings in Eq.~\eqref{eq:L_hvv} and the Wilson coefficients in Eq.~\eqref{eqn:lhelc}.
As a reference, we also compare our conventions to those of the previous SILH
Lagrangian implementation of Ref.~\cite{Alloul:2013naa} and of the
Higgs characterisation Lagrangian of Ref.~\cite{Artoisenet:2013puc}.

\subsection{Constraints from global fits of LEP and LHC Run~I data
\label{ssec:fit}}

In this section we summarise the current bounds on the Wilson coefficients
associated with the effective operators under consideration.

We start from the results of previous works~\cite{Ellis:2014jta,Ellis:2014dva}, 
where a global fit to LEP and LHC Run~I data has been performed. The results 
imply constraints on several linear combinations of the $\bar c$
coefficients appearing in Eq.~\eqref{eqn:lhelc} that we present in
Table~\ref{tab:LHCoperators}, each limit having been obtained by marginalising over all other coefficients. Leading-order (LO) theoretical predictions have been
used and in addition, the modifications of the electroweak
parameters computed in Eq.~\eqref{eqn:e_alpha} and Eq.~\eqref{eqn:cosw} have
not been considered for LHC predictions. We have nevertheless checked that the corresponding effects are
small compared with the LHC Run~I sensitivity, as also noted by the ATLAS
collaboration~\cite{Aad:2015tna}.

\setlength{\tabcolsep}{10pt}
\renewcommand{\arraystretch}{1.6}
\begin{table}
\center
\begin{tabular}{cclc|}
\hline
 Coefficients & Bounds\\
 \hline
$
\frac{\mw^2}{\Lambda^2}(\frac{1}{2}\bar c_{\sss W} -
    \bar c_B)$ &  $[-0.035, 0.005]$ \\ 
  $
  \frac{\mw^2}{\Lambda^2}(\frac{1}{2}\bar c_{\sss W} +\bar c_B)$  &
    $[-0.0033, 0.0018]$ \\ 
  $
  \frac{\mw^2}{\Lambda^2}\,\bar c_{\sss HW}$  & $[-0.07, 0.03]$  \\ 
  $
  \frac{\mw^2}{\Lambda^2}\,\bar c_{\sss HB}$  & $[-0.045, 0.075]$ \\
\hline
\end{tabular}
\caption{
  Current $95\%$ confidence level constraints on the considered effective
  coefficients marginalised in a global fit to LEP and LHC Run~I
  data~\cite{Ellis:2014jta}.}
\label{tab:LHCoperators}
\end{table}
\renewcommand{\arraystretch}{1}

In many classes of SM
extensions (featuring in particular an extended Higgs sector), certain relations
among the coefficients appear. For instance, it is 
common
that matching
conditions such that \mbox{$g^{(2)}_{hww} \propto  \bar c_{HW}+\bar c_{W}  = 0$}
appear~\cite{Gorbahn:2015gxa}. In this case, the global fit
generates the more stringent constraint
\mbox{$\bar c_{HW} = -\bar c_W = [0.0008, 0.04]$} when one sets the effective
scale to \mbox{$\Lambda=\mw$}~\cite{Ellis:2014jta}.

\subsection{Benchmark points \label{ssec:bench}}

For both production processes of interest, we consider two benchmark scenarios in the Wilson coefficient parameter
space. These two points are selected to be compatible with the global fit
results discussed in Sect.~\ref{ssec:fit}.

We first make use of the fact that, as seen in Table~\ref{tab:LHCoperators},
electroweak precision observables strongly constrain a particular linear
combination of the $\bar{c}_W$ and $\bar{c}_B$ Wilson coefficients beyond a
precision than can be hoped for at the LHC.
We therefore impose $\bar{c}_B=-\bar{c}_W/2$, which in turn leads to an allowed
range (setting $\Lambda=\mw$) for $\bar{c}_W$ of $[-0.035,0.005]$, as obtained
from the second constraint on these two parameters.
In order to highlight the impact of the two new Lorentz structures appearing in
the interaction vertices of the Lagrangian of Eq.~\eqref{eq:L_hvv}, we allow for
non-zero values for both the $\bar{c}_{HW}$ and $\bar{c}_W$ coefficients.

Our benchmark scenarios are defined in Table~\ref{tab:bench}. In the first
setup, we only switch on the ${\cal O}_{\sss HW}$ operator (which induces new
physics contributions to both the $g^{(1)}_{hvv}$ and $g^{(2)}_{hvv}$
structures). With the second point, we additionally fix $\bar{c}_{W}$ to an
equal and opposite value relying on the constraint relation brought up in
Sect.~\ref{ssec:fit}. This allows for turning on solely the $g^{(1)}_{hvv}$
coupling (see Table~\ref{tab:L3conv}).

 \begin{table}
	 \center
     \begin{tabular}{l|ccccc}
         \hline
         Benchmarks & $\bar{c}_{HW}$ & $\bar{c}_{W}$ & $\bar{c}_{B}$ 
		 \tabularnewline
         \hline
         A: $g^{(1)}_{hvv},g^{(2)}_{hvv}\ne0$ & 0.03 & 0 & 0 \tabularnewline
         B: $g^{(1)}_{hvv}\ne0,\,g^{(2)}_{hvv}=0 $ & 0.03 & $-0.03$ & 0.015 \tabularnewline
         \hline
     \end{tabular}
         \caption{EFT benchmark points under consideration, with $\Lambda=\mw$.
          We additionally set $\bar{c}_{BB}=\bar{c}_{HB}=0$ for simplicity.
\label{tab:bench}} 
\end{table}

\section{Setup for NLO+PS simulations\label{sec:sim}}

Our numerical results are derived at the NLO accuracy in QCD thanks to a joint
use of the {\sc FeynRules}/{\sc NloCT} and {\sc MG5\_aMC} packages. The EFT
Lagrangian of Eq.~\eqref{eqn:lhelc} has been implemented in
{\sc FeynRules}~\cite{Alloul:2013bka}, while the computation of the ultraviolet
counterterms and the rational $R_2$ terms necessary for numerical
loop-integral evaluation has been done by {\sc NloCT}~\cite{Degrande:2014vpa}
that relies on {\sc FeynArts}~\cite{Hahn:2000kx}. The model information is
then provided to {\sc MG5\_aMC}~\cite{Alwall:2014hca} in the UFO
format~\cite{Degrande:2011ua}.
Within {\sc MG5\_aMC}, loop-diagram contributions are numerically
evaluated~\cite{Hirschi:2011pa} and combined with the real emission pieces
within the FKS subtraction scheme~\cite{Frixione:1995ms,Frederix:2009yq}.
Short-distance events are finally matched to parton showers according to the
MC@NLO prescription~\cite{Frixione:2002ik}.

We generate events for 13~TeV LHC collisions using the LO and NLO NNPDF2.3 set
of parton densities~\cite{Ball:2013hta} for LO and NLO simulations, respectively.
Events are then showered and hadronised within the {\sc Pythia8}
infrastructure~\cite{Sjostrand:2007gs}, which is also used for handling
Higgs-boson decays. This latter step relies on
{\sc eHdecay}~\cite{Contino:2014aaa}
that computes all branching fractions of the Higgs boson into the relevant final
states to the first order in the Wilson coefficients. This procedure has the
advantage of providing a correct normalisation for the production rates that
includes all effects originating from the EFT operators. For the two adopted
benchmark points, the deviations from the SM branching ratios are found to be very small.

Event reconstruction and analysis are performed using the
{\sc MadAnalysis5}~\cite{Conte:2012fm} framework, which makes use of all jet
algorithms implemented in the {\sc FastJet} program~\cite{Cacciari:2011ma}.
Jets are defined using the anti-$k_T$ algorithm~\cite{Cacciari:2008gp} with a
radius parameter of 0.4.

Theoretical uncertainties due to renormalisation ($\mu_R$) and factorisation
($\mu_F$) scale variations are accounted for thanks to the reweighting
features of {\sc MG5\_aMC}~\cite{Frederix:2011ss}. At the event generation
stage, nine alternative weights are stored for each event, corresponding to the
independent variation of the two scales by a factor of two up and down with
respect to a central scale $\mu_0$.
Since the parton shower is unitary, these could be used to reweight the events
after showering and reconstruction, saving a great deal of computational time and storage.
The scale variation uncertainty is taken to be the largest difference between the central scale and the alternative scale choice predictions.
We use as a central scale $\mu_0=H_T/2$ and $m_W$ for VH and VBF processes,
respectively,
where $H_T$ is defined at the parton-level as the scalar sum of the transverse
momentum of all visible final-state particles and the missing transverse energy.
We refer the reader to the appendix for further technical details on event generation.

\section{Higgs production in association with a vector boson\label{sec:WH}}

Higgs-boson production in association with a vector boson is an excellent probe
for new physics, as the momentum transfer in the process is directly sensitive to
the Lorentz structure appearing in the interaction vertices~\cite{Ellis:2012xd}.
The use of differential
information at the LHC Run~I has therefore enhanced the sensitivity of Higgs
data to possible new physics effects~\cite{Ellis:2014dva}. Those Run~I studies
have, however, relied on predictions evaluated at the LO accuracy in QCD. With
the improved capabilities of the LHC Run~II, NLO QCD effects become more
relevant and more precise predictions are in order.

To showcase our NLO simulation setup for associated VH production, we study
various differential distributions in the
\begin{equation}
p\, p \to H \, W^+ \to b \, \bar b \, \ell^+ + \slashed E
\end{equation}
channel, where $\slashed E$ stands for the final-state missing energy. We
impose the requirement that both $b$-jets and leptons have a pseudorapidity, $\eta$, and
a transverse momentum, $p_T$, satisfying \mbox{$|\eta|<2.5$} and
\mbox{$p_T>25$~GeV}, respectively, while non-$b$-tag\-ged jets are instead
allowed
to be more forward, with \mbox{$|\eta|<4$}, for the same $p_T$ requirement. We
select events by demanding the presence of one lepton and two $b$-jets based on
truth-level hadronic information, a $b$-tagged jet being defined by the presence
of a $b$-hadron within a cone of radius $R=0.4$ centred on the jet direction.

\begin{figure*}
\center
\includegraphics[width=0.325\textwidth]{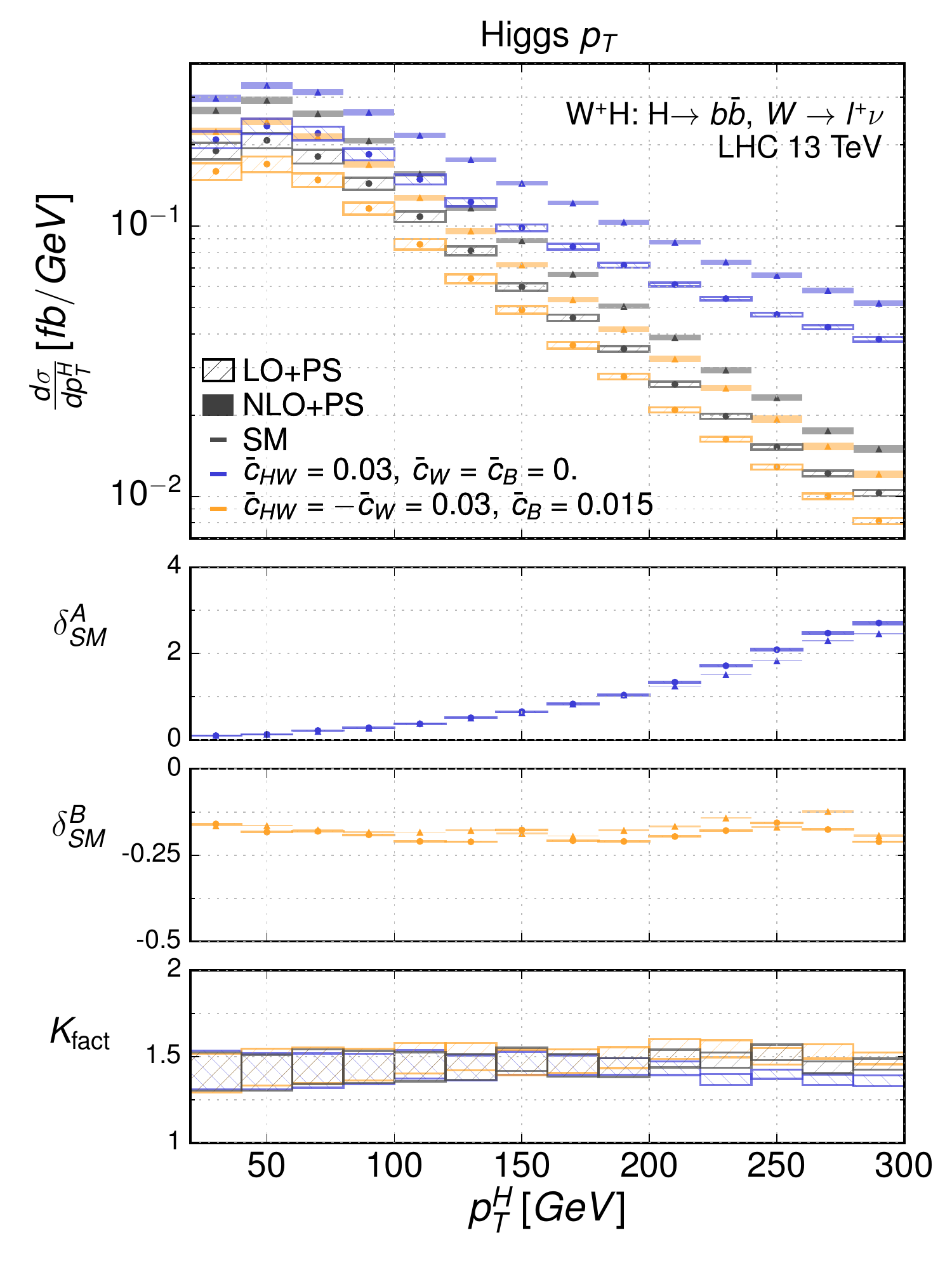}
\includegraphics[width=0.325\textwidth]{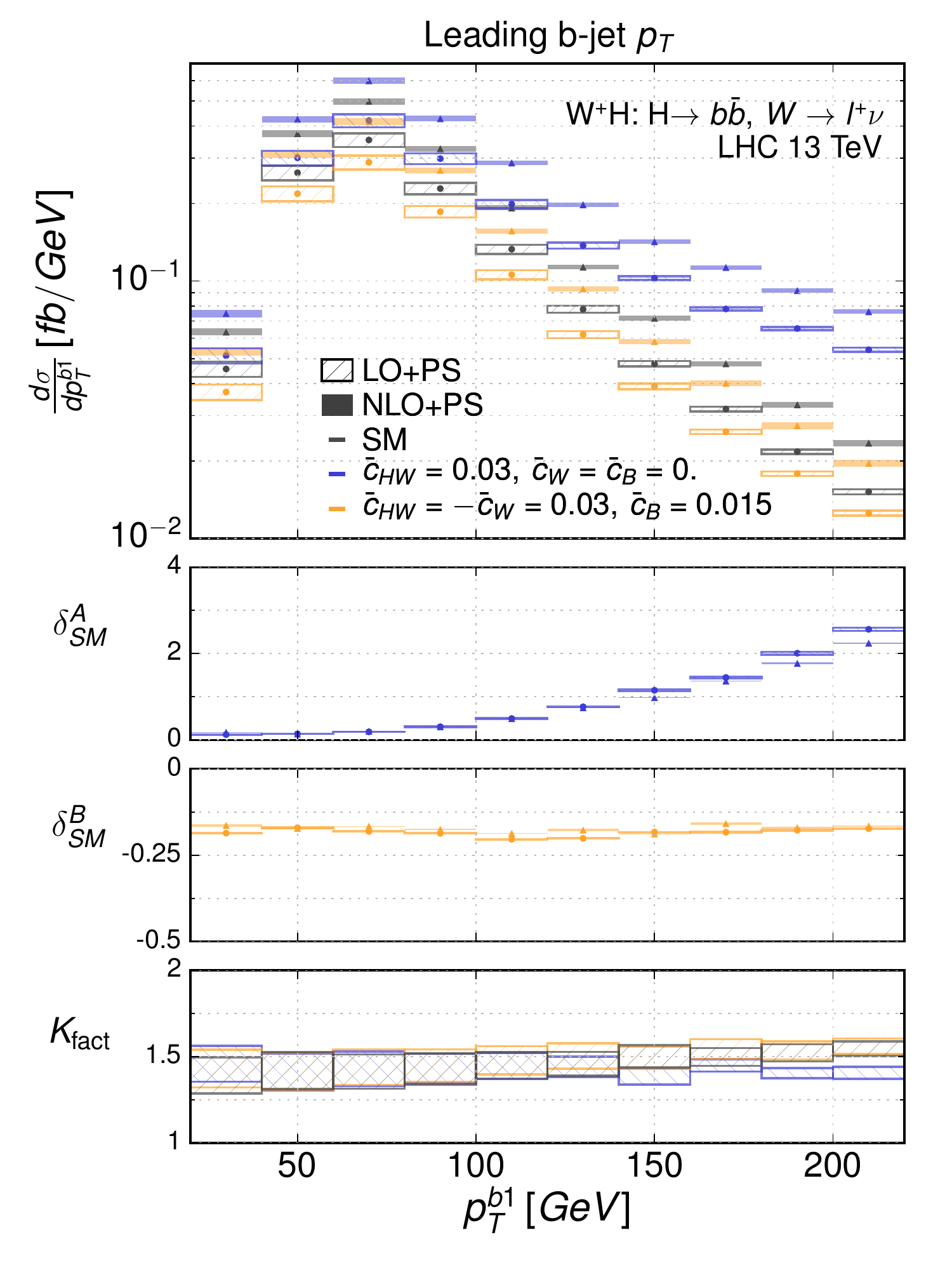}
\includegraphics[width=0.325\textwidth]{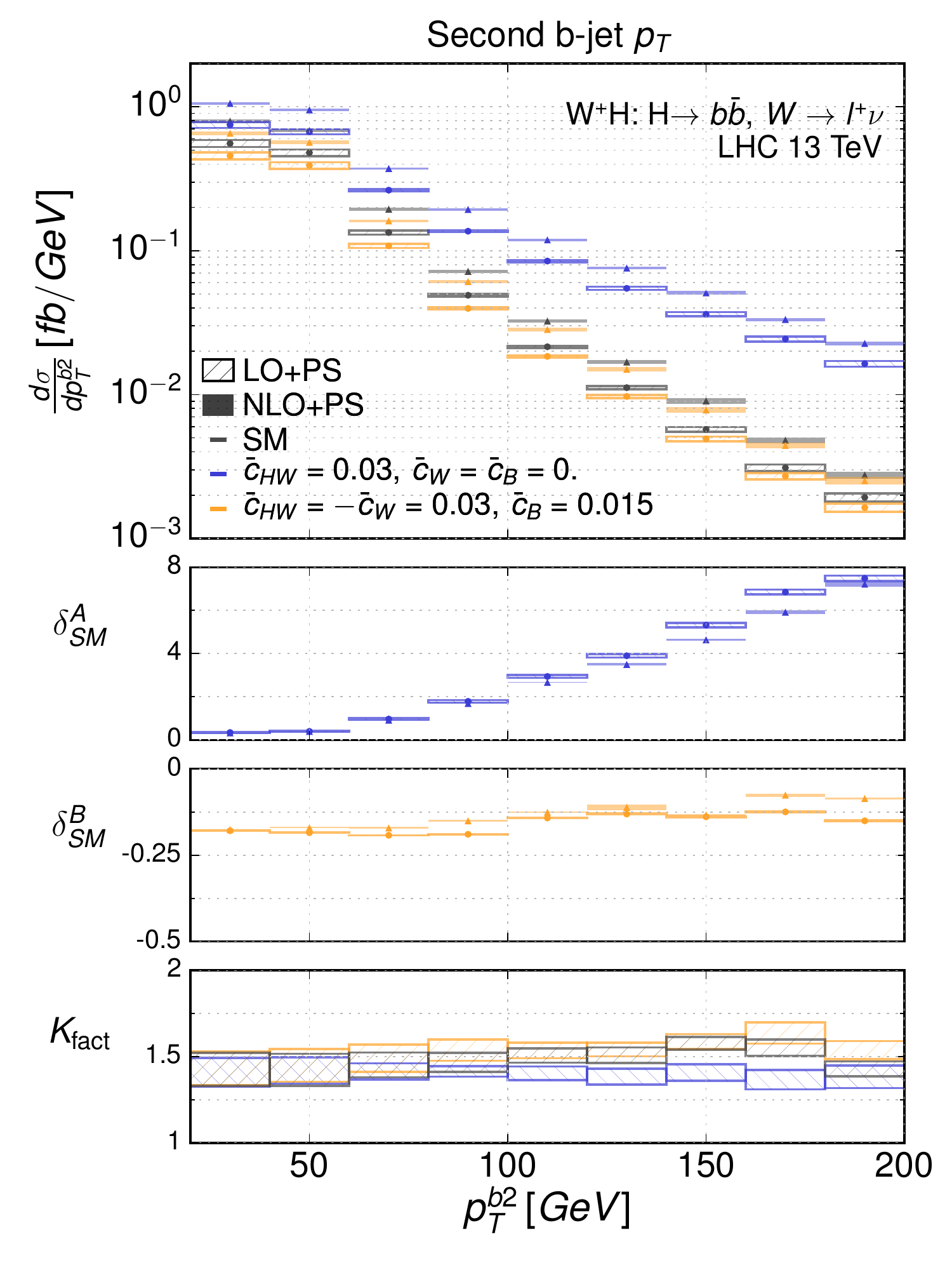}
\includegraphics[width=0.325\textwidth]{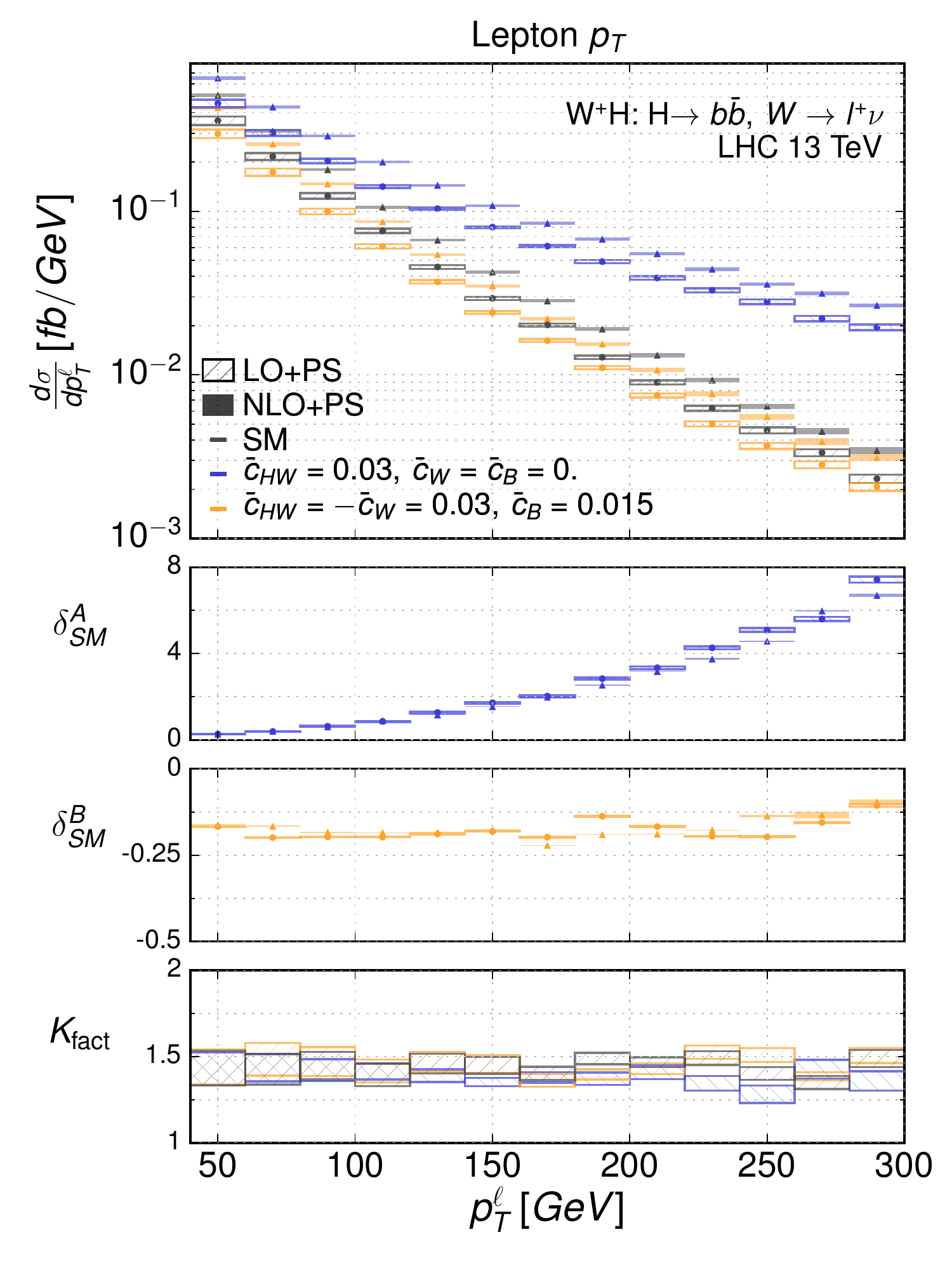}
\includegraphics[width=0.325\textwidth]{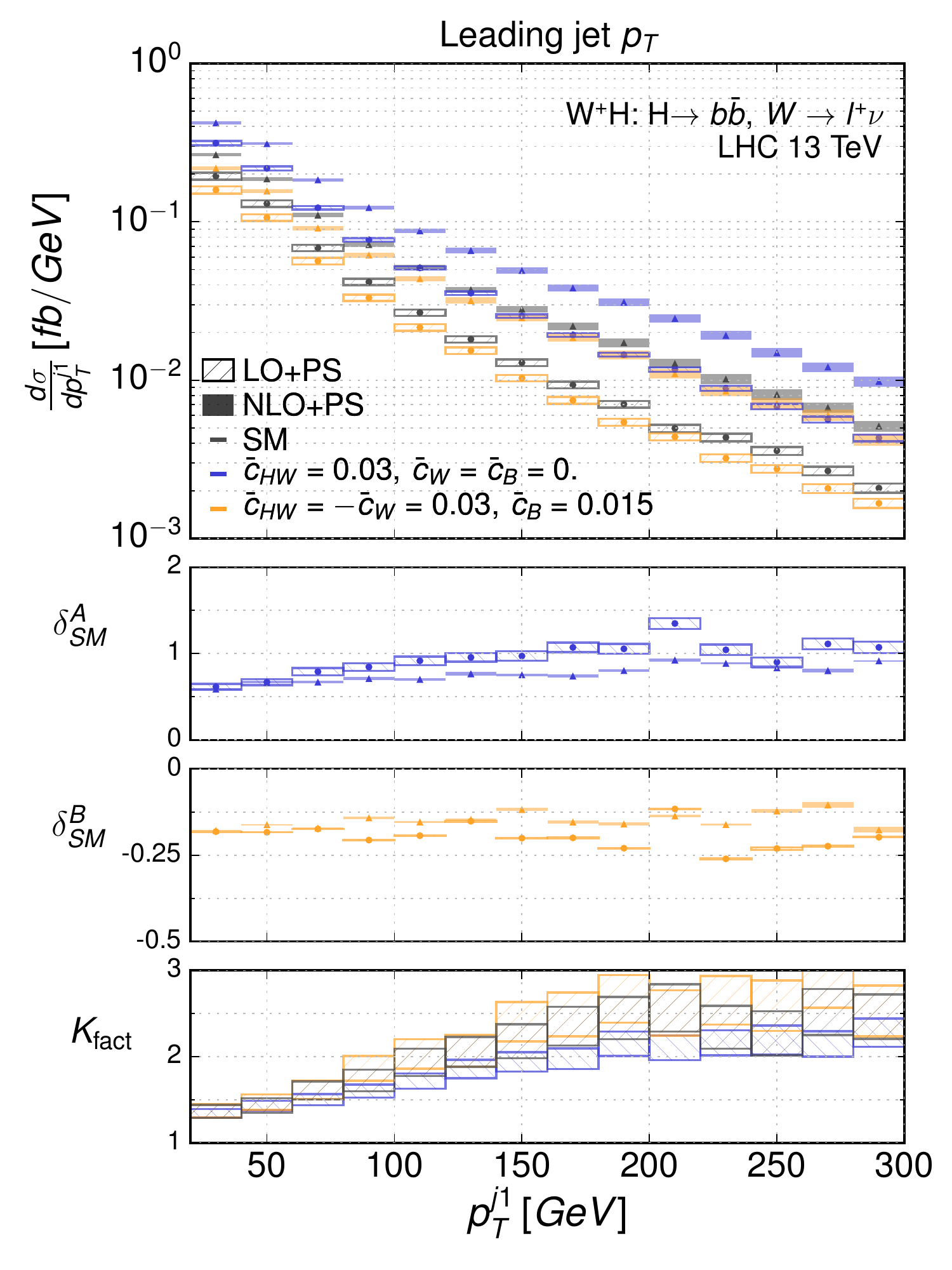}
\includegraphics[width=0.325\textwidth]{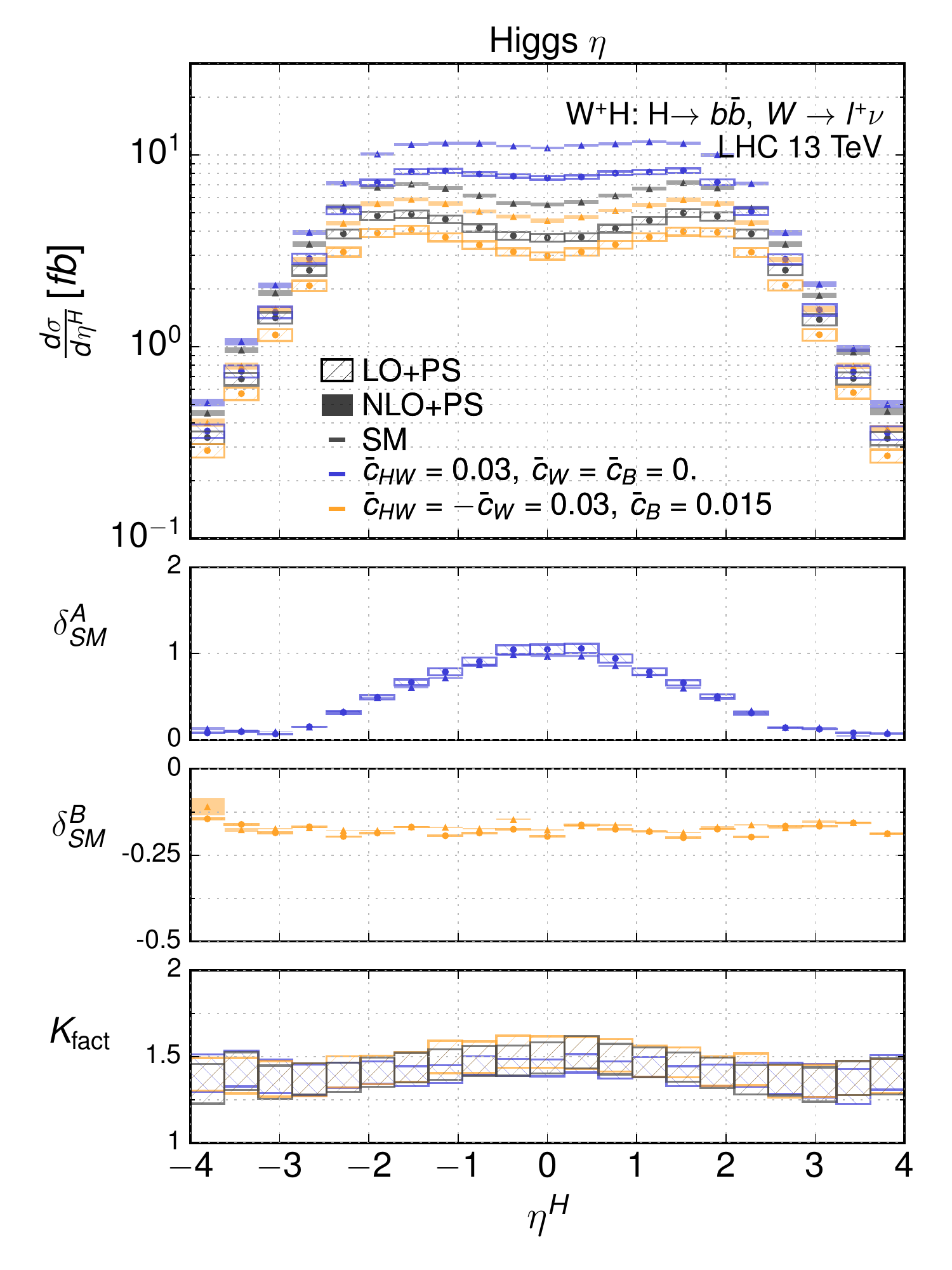}
\includegraphics[width=0.325\textwidth]{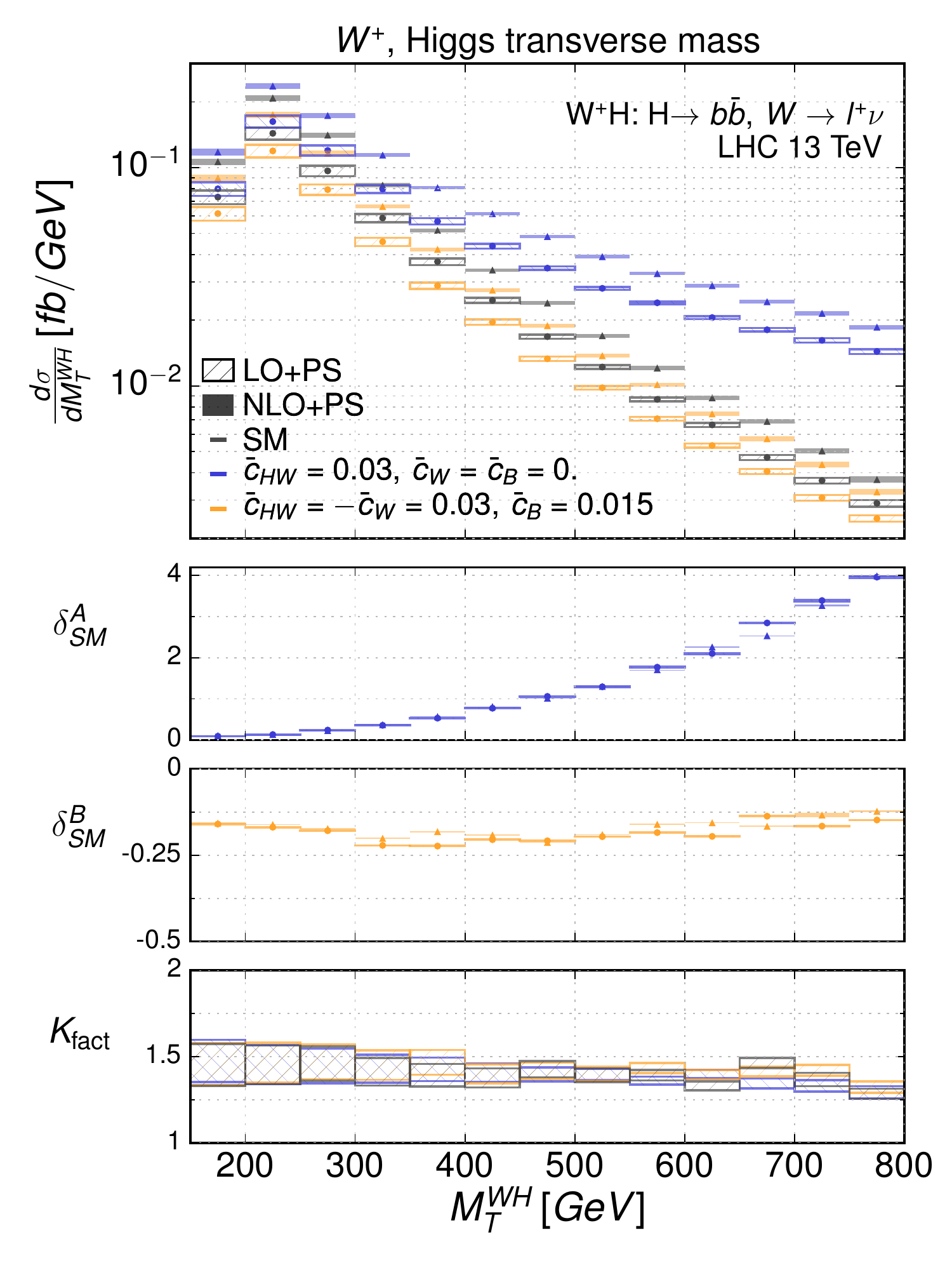}
\includegraphics[width=0.325\textwidth]{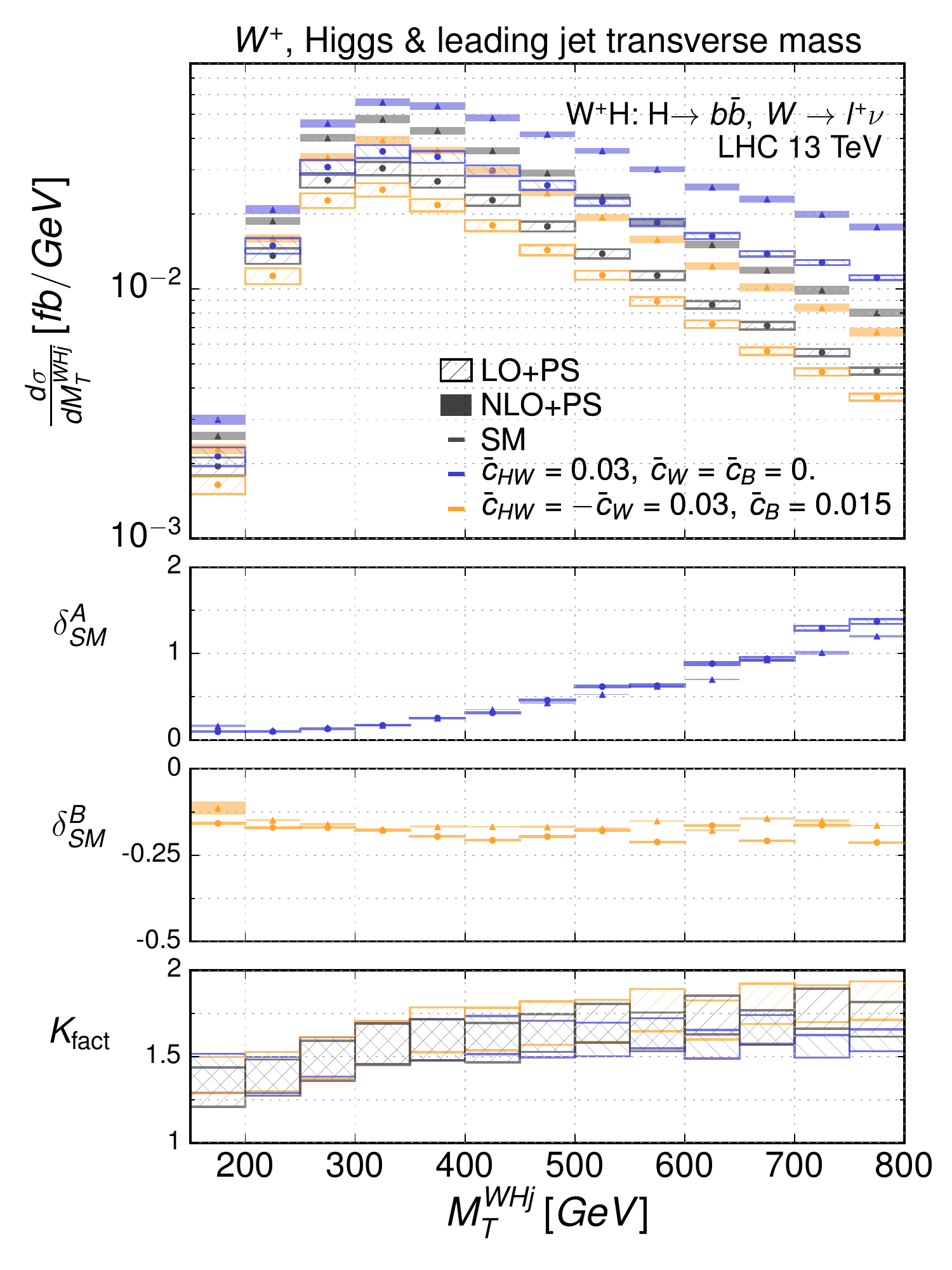}
\includegraphics[width=0.325\textwidth]{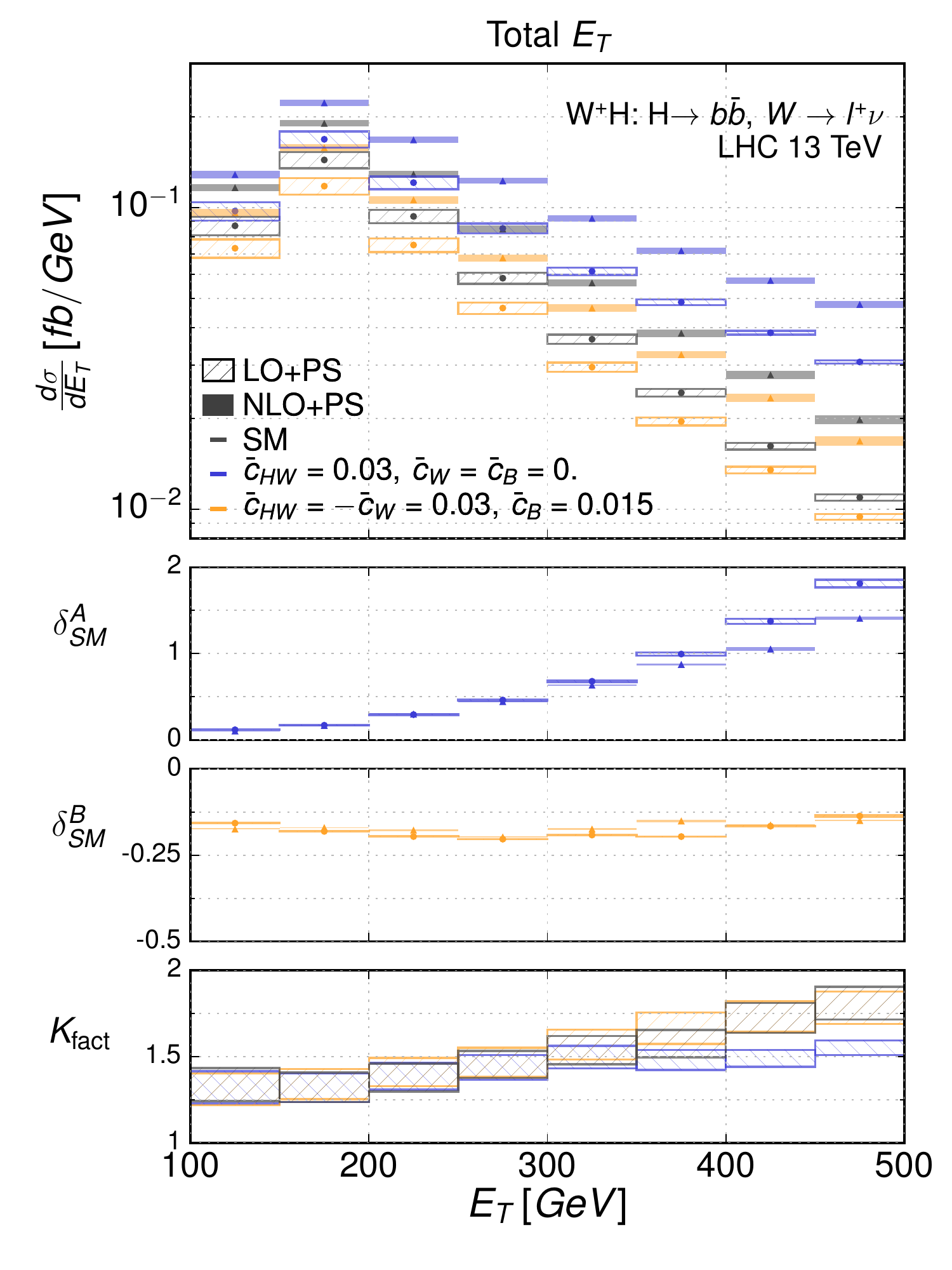}
\caption{Selection of $WH$ differential distributions at the (N)LO+PS accuracy
 in which the SM predictions are compared with results obtained for the two benchmark scenarios discussed in Sect.~\ref{ssec:bench}. Our
 predictions include the theoretical uncertainties stemming from scale
variation. \label{fig:WH}}
\end{figure*}

In Fig.~\ref{fig:WH}, we present the transverse momentum spectrum of the
$b\bar b$ system (upper left), of the leading (upper centre) and
next-to-leading (upper right) $b$-jets, of the lepton (middle left) and of
the leading jet (middle centre). We then focus on the distribution in
pseudorapidity for the $b\bar b$ system (middle right), in the transverse mass
of the $W$-boson and Higgs boson (lower left) and of the $W$-boson, Higgs boson
and leading-jet system (lower centre) and in the total transverse energy (lower right). 
In each subfigure, the results are shown both at the LO+PS and NLO+PS 
accuracies, together with uncertainties related to scale variation.

\begin{figure*}
\center
\includegraphics[width=0.325\textwidth]{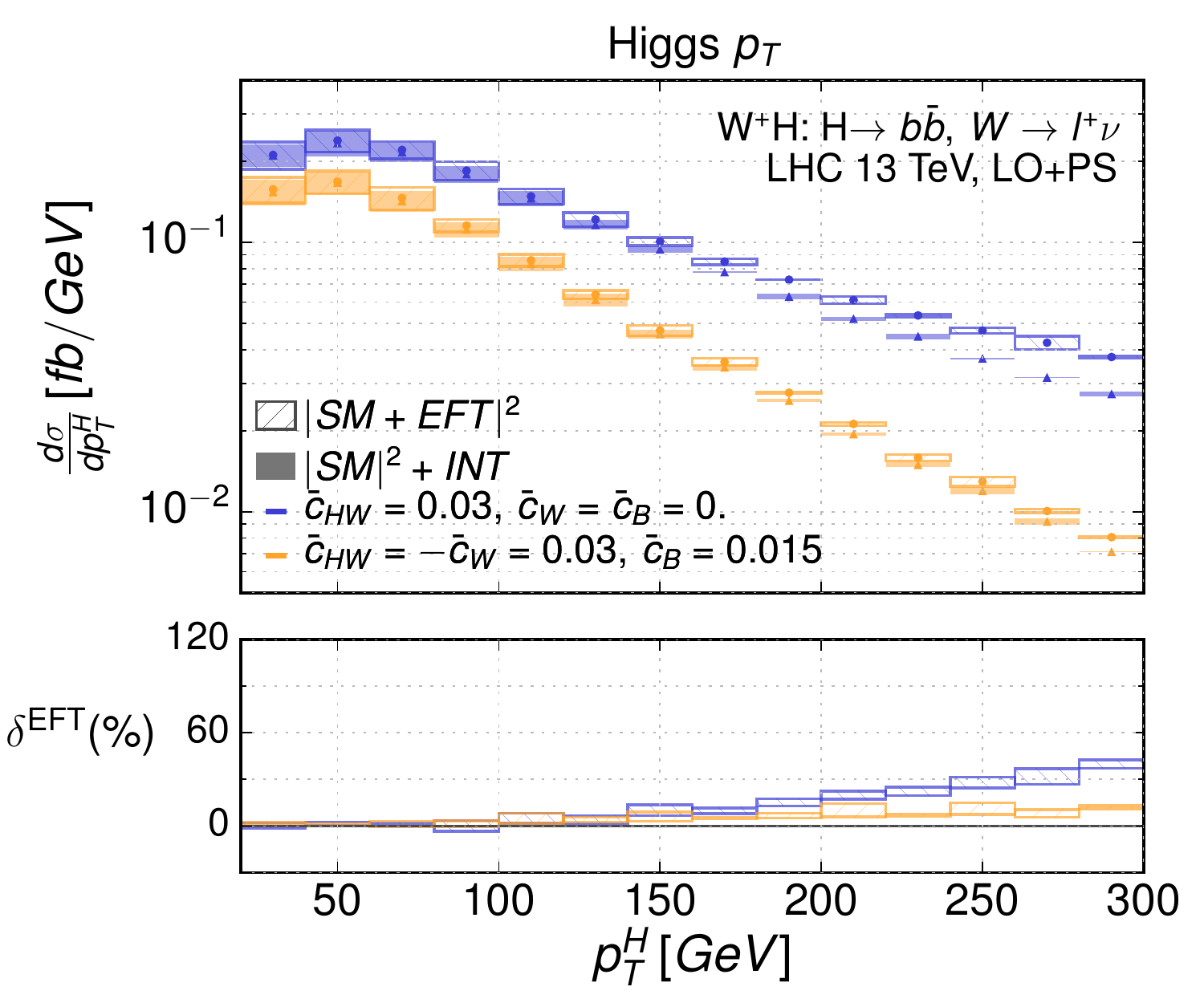}
\includegraphics[width=0.325\textwidth]{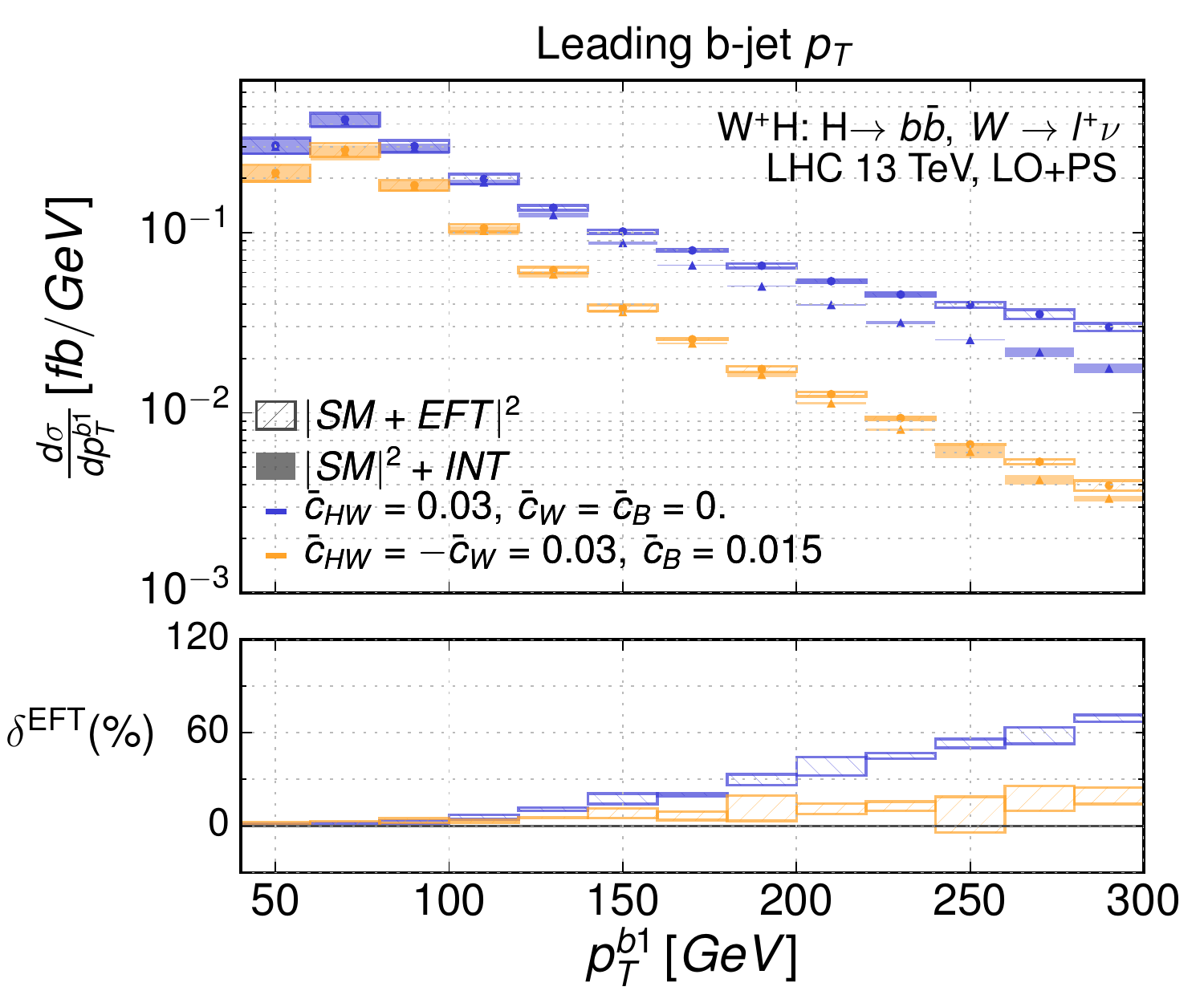}
\includegraphics[width=0.325\textwidth]{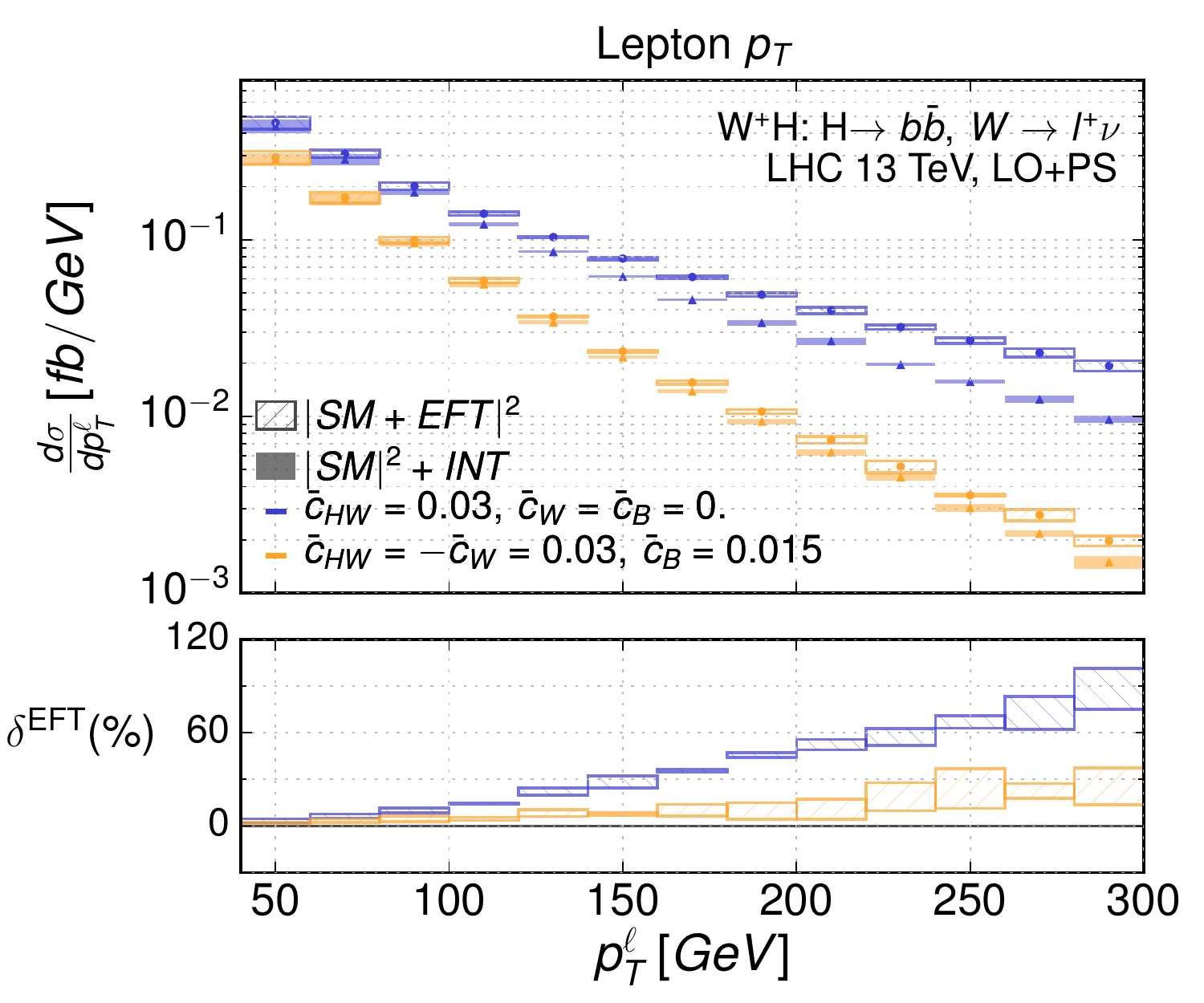}
  \caption{Selection of $WH$ differential distributions at the LO+PS accuracy
  in which the EFT squared contributions are either included or not. The results
  are presented for both benchmark scenarios described in
  Sect.~\ref{ssec:bench}. \label{fig:WH_INT}}
\end{figure*}

For each studied observable, we investigate in the first two lower bands of each
subfigure the relative difference between the predictions in the SM and in the
case of both considered benchmark points A and B,
\begin{align}
    \delta_{SM}^i =  \frac{\sigma_{i}}{\sigma_{SM}}-1\qquad\text{for }\quad i=A,B\ .
\end{align}
In the last band, we additionally show differential $K$-factors defined as the
binned ratio of NLO to LO predictions taking only the total NLO uncertainty into account.

The predictions are found to be stable under radiative corrections, as expected
for any process with a Drell--Yan-like topology. The obtained $K$-factors are
indeed relatively flat and independent of the EFT parameters, with the exception
of the observables that rely on the leading-jet kinematics which turn out to be
much harder at NLO. Hard QCD radiation contributions originating from the matrix
element are in this case included, in contrast to the LO setup where QCD
radiation is only described by the parton shower and thus modelled in the
soft-collinear kinematical limit.

LO predictions are found to be inaccurate and do not overlap with the NLO results
even after considering scale variation uncertainties. This is particularly true
at high transverse momentum $p_T$, transverse mass $M_T$ and total transverse
energy $E_T$. This behaviour is once again expected for a
Drell--Yan-like process that does not depend on $\alpha_S$ at fixed LO. If one
were to use the difference between the LO and NLO results as an error estimate
for the LO predictions and the scale variation only for the NLO, then the
reduction of the theory error would be better reflected  
by Fig.~\ref{fig:WH}. The $\delta^i_{SM}$ ratios also
remain stable with respect to QCD corrections, except at very high energies for
the benchmark point A where small differences appear between the LO and NLO
predictions. These would, by construction, be covered by the aforementioned
improved definition of the LO theoretical uncertainties.

All distributions strongly depend on the value of the EFT Wilson coefficients.
For the adopted scenario A, large enhancements are observed in the tails of the
$p_T$, $M_T$ and $E_T$ distributions, which correspond to a centrally produced
$b\bar b$ system (with a small pseudorapidity). In contrast, event rates are
only
rescaled by about \mbox{15--20\%} with respect to the SM for the scenario B. This
originates from the $g^{(2)}_{hvv}$ coupling that vanishes in this scenario, so
that only the $g^{(1)}_{hvv}$ coupling drives the EFT behaviour in the
high-energy tails. However, this latter coupling is known to yield a smaller
impact than the $g^{(2)}_{hvv}$ coupling~\cite{Maltoni:2013sma,Mimasu:2015nqa},
and it is therefore the presence of the $g^{(2)}_{hvv}$ interaction vertex in
scenario A that leads to the large observed deviations. This
constitutes a very promising avenue for setting limits on EFT parameters
from $WH$ studies and similar behaviour can be observed for $ZH$ production,
where the $\bar{c}_{HB}$ and $\bar{c}_{BB}$ coefficients additionally play a
role. In this case, the gluon fusion initiated contribution should however also be considered, as
discussed in Refs.~\cite{Mimasu:2015nqa,Bylund:2016phk}.

While such large enhancements can be exploited to obtain powerful constraints on
the SMEFT Wilson coefficients, they do raise the question of the validity of the
EFT approach at large momentum transfer~\cite{Ellis:2014dva,Biekoetter:2014jwa,Englert:2014cva,Contino:2016jqw}. This question
could be addressed with the use of dedicated benchmark models to compare the
breakdown of the EFT framework against well-motivated ultraviolet-complete
models~\cite{Gorbahn:2015gxa,Brehmer:2015rna}. At a more simplistic level we
can also make use of the {\sc MG5\_aMC} ability to select only interference
contributions (at LO) to assess the impact of the squared EFT terms given our benchmark
choices (technical details are described in Appendix~\ref{sec:implementation}).
Fig.~\ref{fig:WH_INT} shows a selection of distributions, overlaying
predictions with and without this squared term. We observe significant 
differences between the two choices that are greater than the scale uncertainty of the 
predictions. Depending on the observable, these can range from 40 to 100\% 
on the interference-only prediction for  the benchmark scenario A, while they are much milder for the
benchmark scenario B. This suggests that current sensitivities on this region of the Wilson 
coefficient parameter space may not yet lend themselves to an EFT interpretation within the 
validity of the framework. A reduction of the production rate from the SM 
value, as seen for benchmark scenario B, moreover indicates the dominance of 
the interference term between the SM and EFT contributions given that the 
squared terms are positive-definite (Fig.~\ref{fig:VBF}). 

%

\section{Higgs production via vector boson fusion\label{sec:VBF}}

\begin{figure*}
\center
 \includegraphics[width=0.325\textwidth]{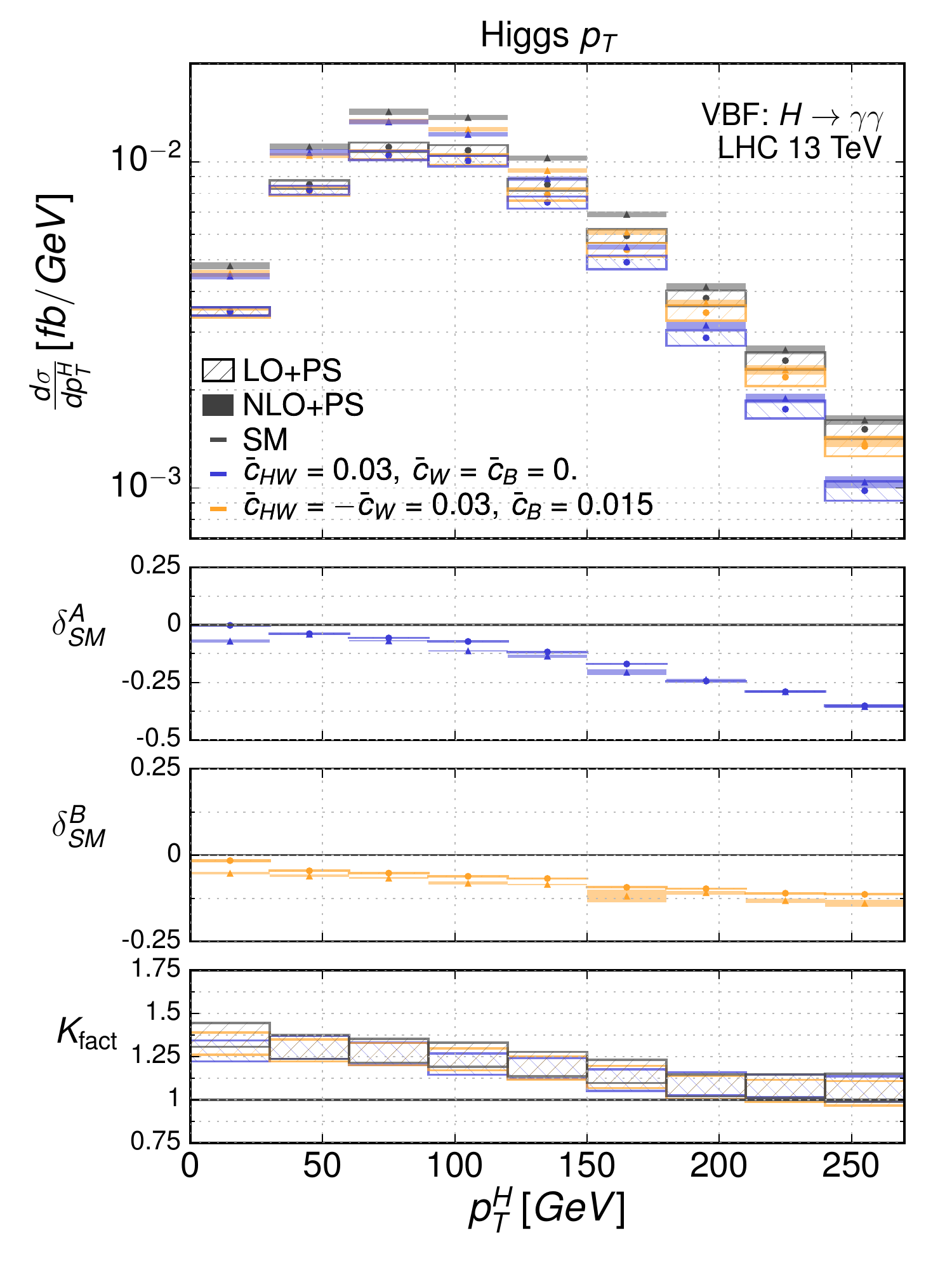}
 \includegraphics[width=0.325\textwidth]{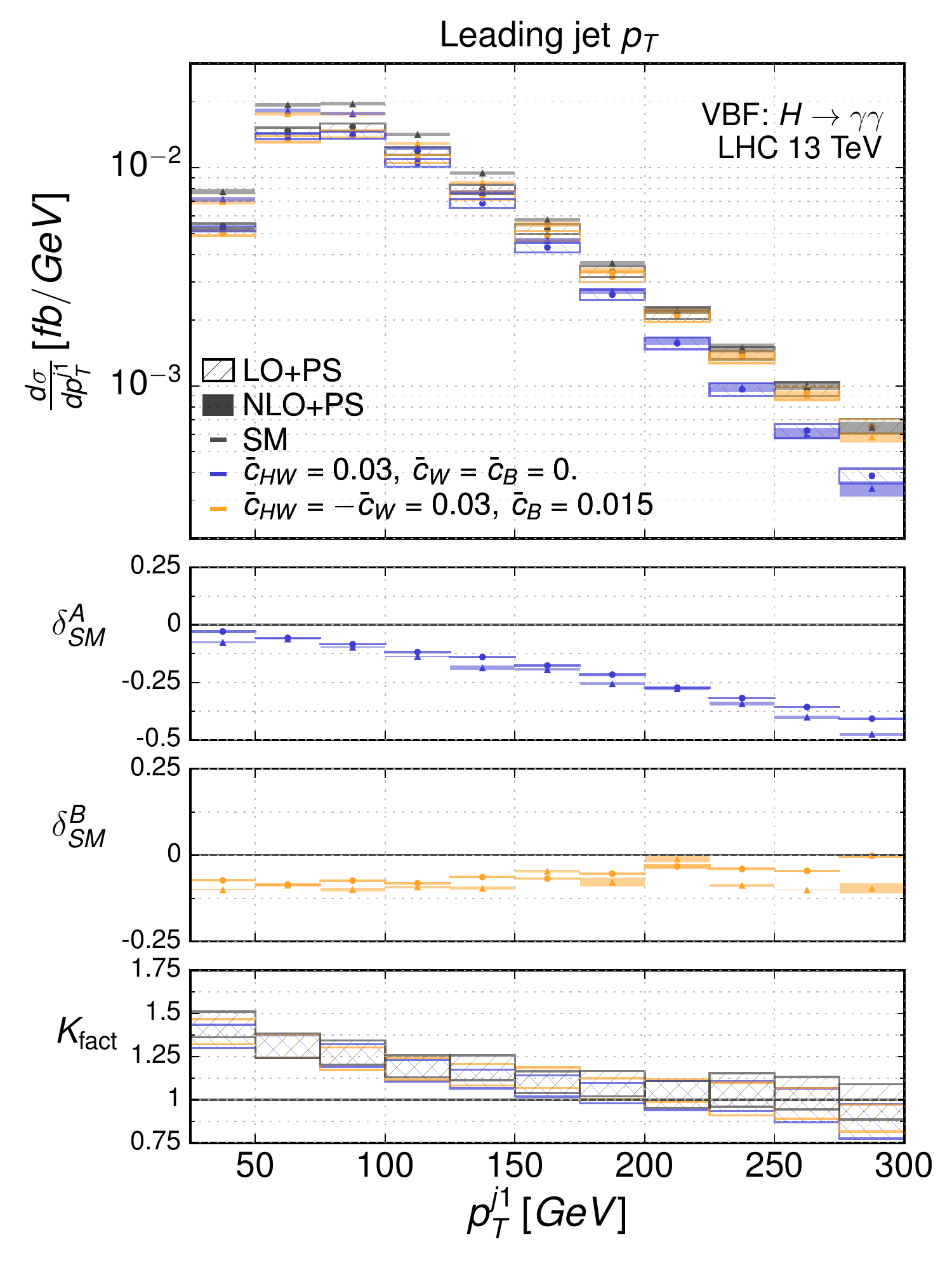}
 \includegraphics[width=0.325\textwidth]{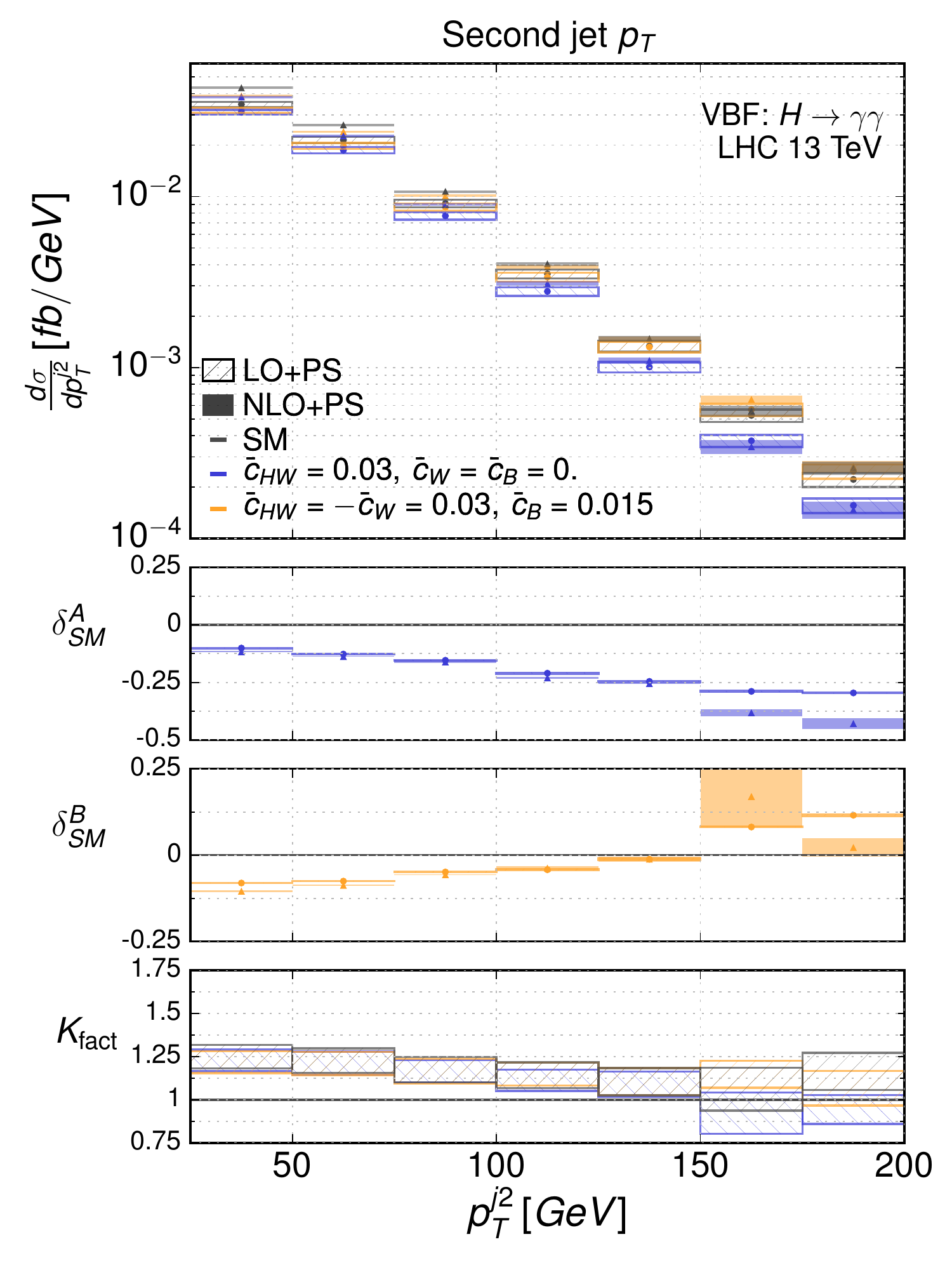}
 \includegraphics[width=0.325\textwidth]{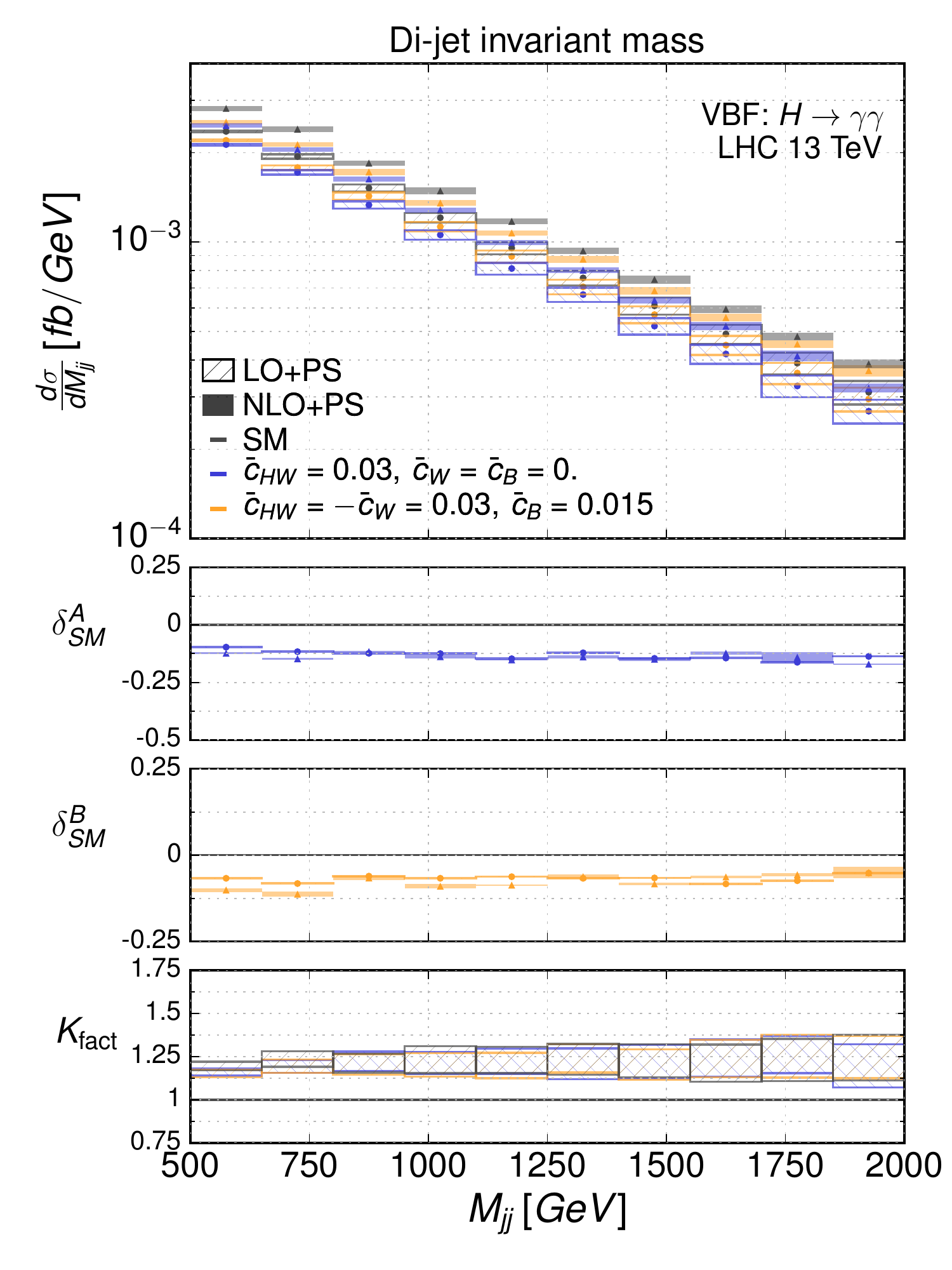}
 \includegraphics[width=0.325\textwidth]{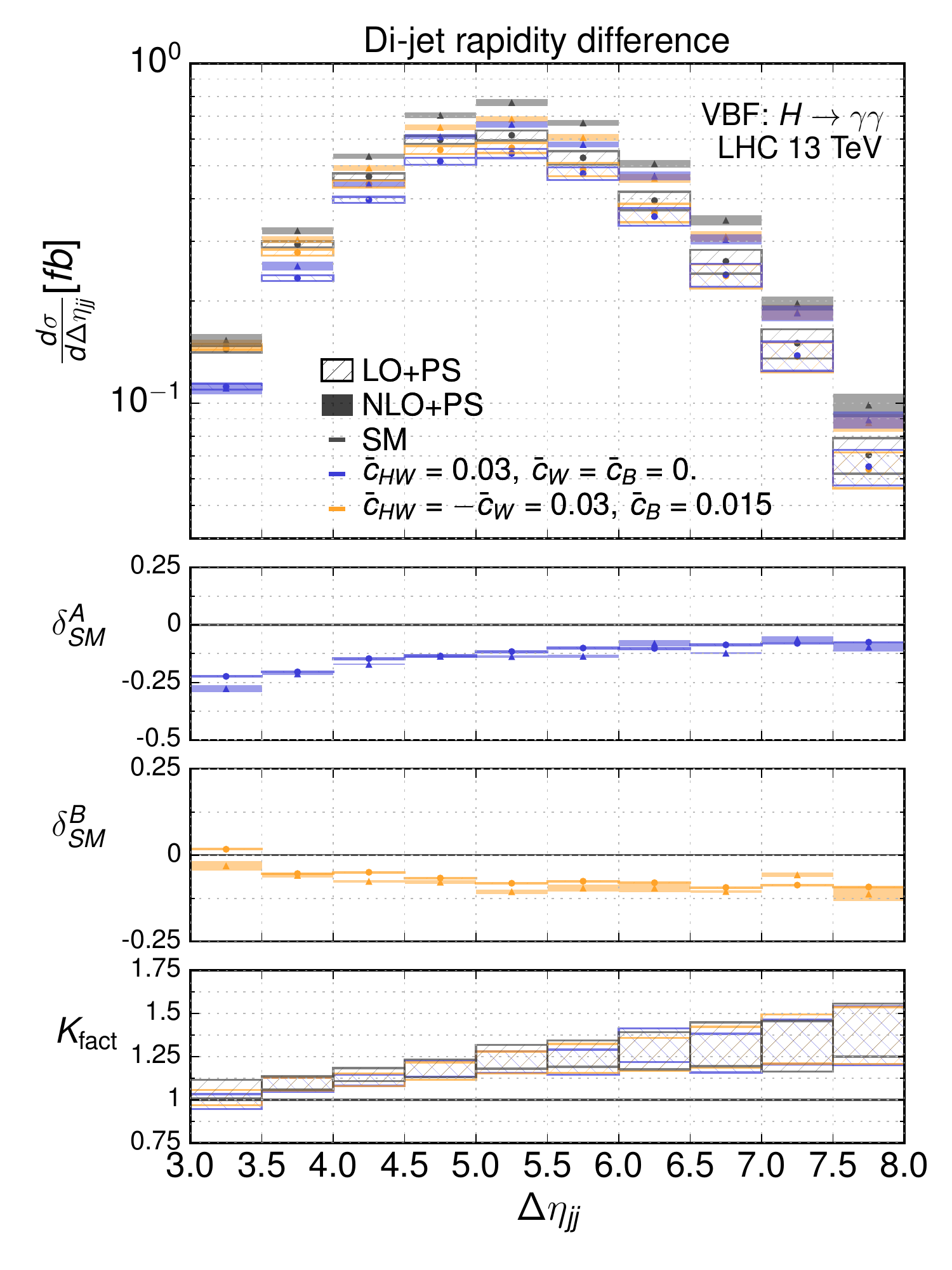}
 \includegraphics[width=0.325\textwidth]{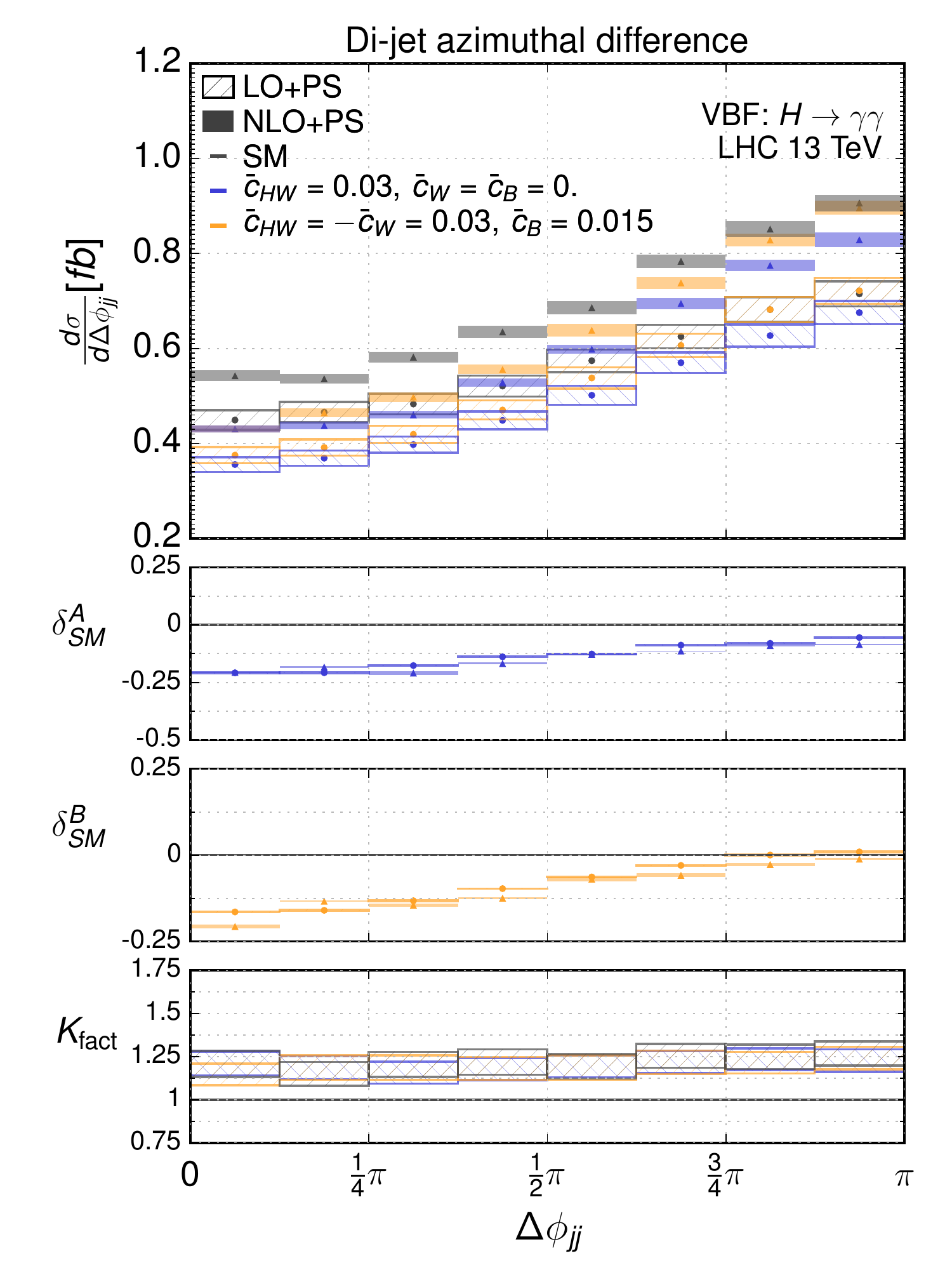}
 \caption{Same as in Fig.~\ref{fig:WH}, but for VBF Higgs production.
 \label{fig:VBF}}
\end{figure*}

Another powerful probe of anomalous higher-derivative interactions between
weak and Higgs bosons consists of the VBF Higgs production mode where it is
produced in association with two forward jets,
\begin{equation}
 p\,p \to (H \to \gamma\,\gamma) + j\,j \ .
\end{equation}
Our event selection requires the presence of at least two jets with a
pseudorapidity \mbox{$|\eta|< 4.5$} and a transverse momentum \mbox{$p_T>25$~GeV}, and we additionally impose the requirement that the Higgs boson
decays into a pair of photons with a pseudorapidity satisfying
\mbox{$|\eta|<2.5$} and a transverse momentum \mbox{$p_T>20$~GeV}.
We moreover include a standard VBF selection on the invariant
mass $M_{jj}$ and pseudorapidity separation $\Delta\eta_{jj}$ of the pair of
forward jets,
\begin{align}
 M_{jj} >500~{\rm GeV}\quad {\rm and}\quad  \Delta \eta_{jj} >3 \ .
\end{align}

Several kinematical observables are sensitive to the momentum flow in the VBF
process, for which EFT contributions deviate from the SM prediction. We consider in
Fig.~\ref{fig:VBF} the distribution in the transverse momentum of the diphoton
system (upper left), in the $p_T$ of the leading (upper centre) and subleading
(upper right) jets, in the invariant mass of the dijet system (lower left), as
well as in its pseudorapidity (lower centre) and azimuthal angular (lower right)
separations. 
The consistent definition of scale uncertainties that are possible with the NLO predictions helps to quantify the discriminatory power between the new physics benchmarks and the SM.
Similarly to the VH process, the NLO corrections are independent of the EFT
parameters and cannot be completely described by an overall $K$-factor.

In contrast to the VH process, we observe a depletion of the production rate for
both benchmark scenarios, which mainly impacts the high-energy tails of the
differential distributions. This indicates that our predictions may be safer with respect to the validity of the EFT, as it implies that the interference term dominates over the EFT
squared one. In particular, the effects for the benchmark scenario B are more pronounced with respect to the SM compared to the $VH$ production case and show some different shape deformations. This illustrates the complementarity between the VH and
VBF processes in disentangling the possible EFT sources for any potential deviation. Although the
correlations between the forward jets as well as between the jets and the Higgs
boson are known to be sensitive to new physics effects~\cite{Englert:2012xt,%
Djouadi:2013yb}, those are less sensitive than the individual Higgs and jet
$p_T$ distributions for our two benchmark scenarios.

\begin{figure*}[h!]
\center
\includegraphics[width=0.325\textwidth]{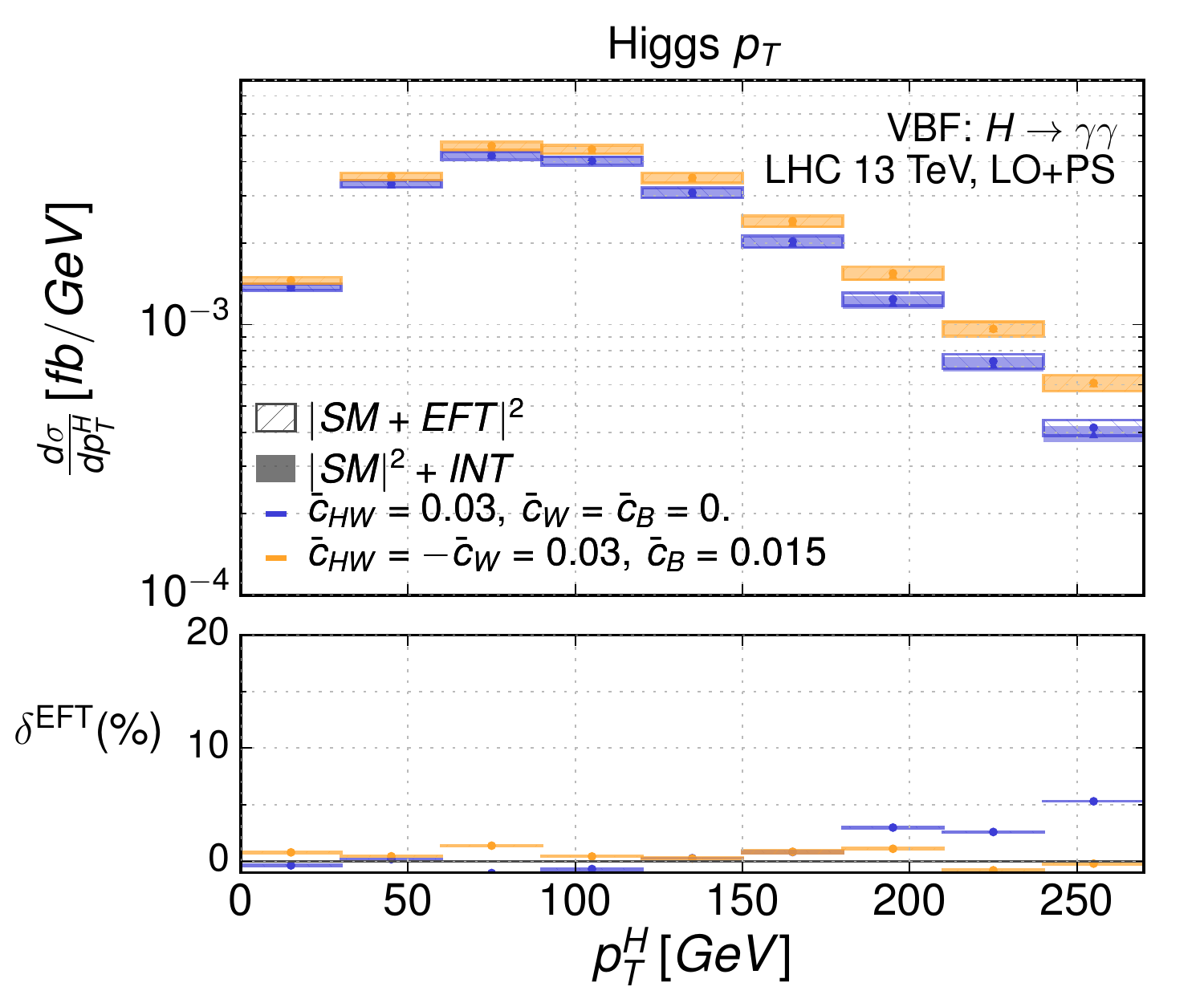}
\includegraphics[width=0.325\textwidth]{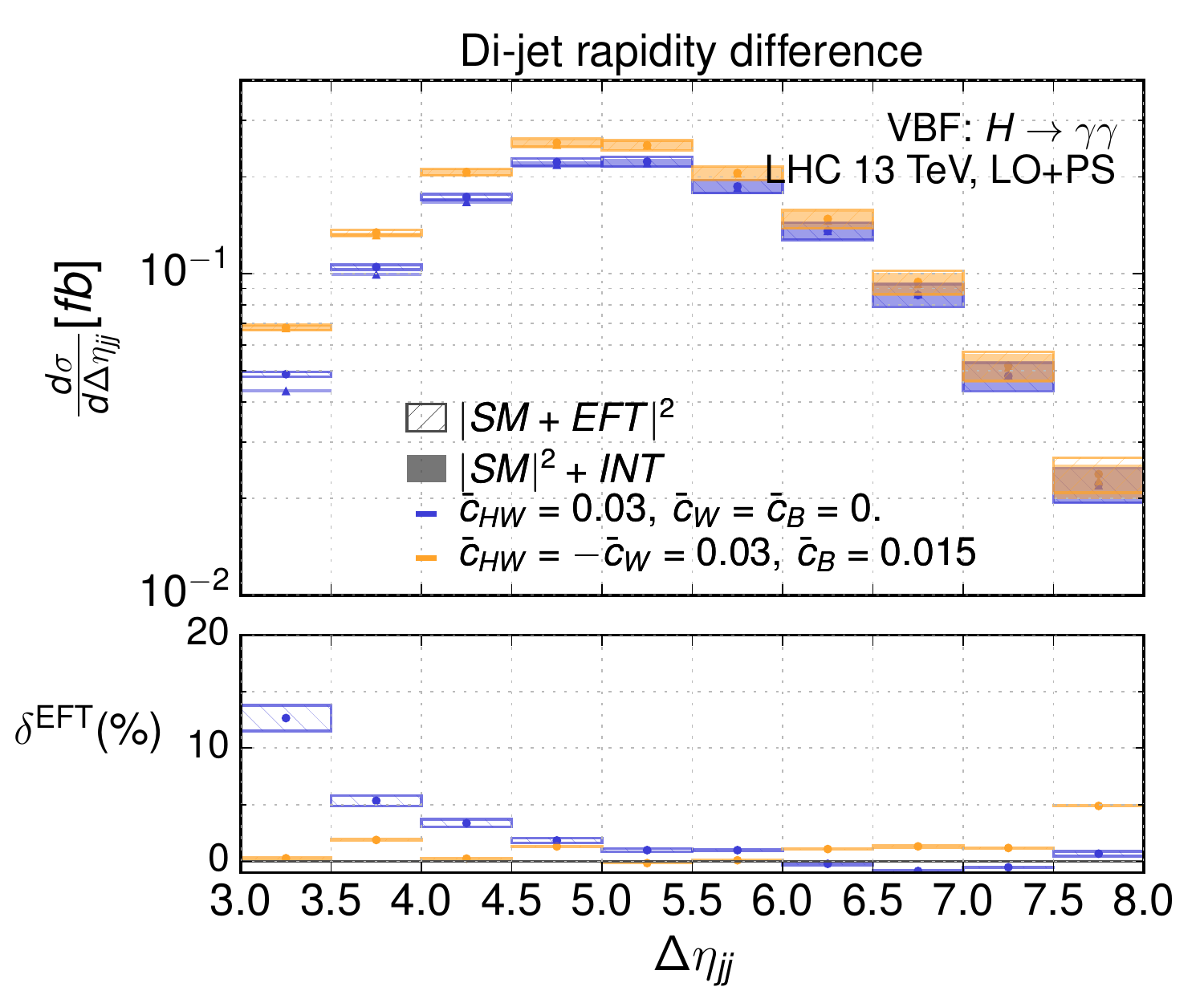}
\includegraphics[width=0.325\textwidth]{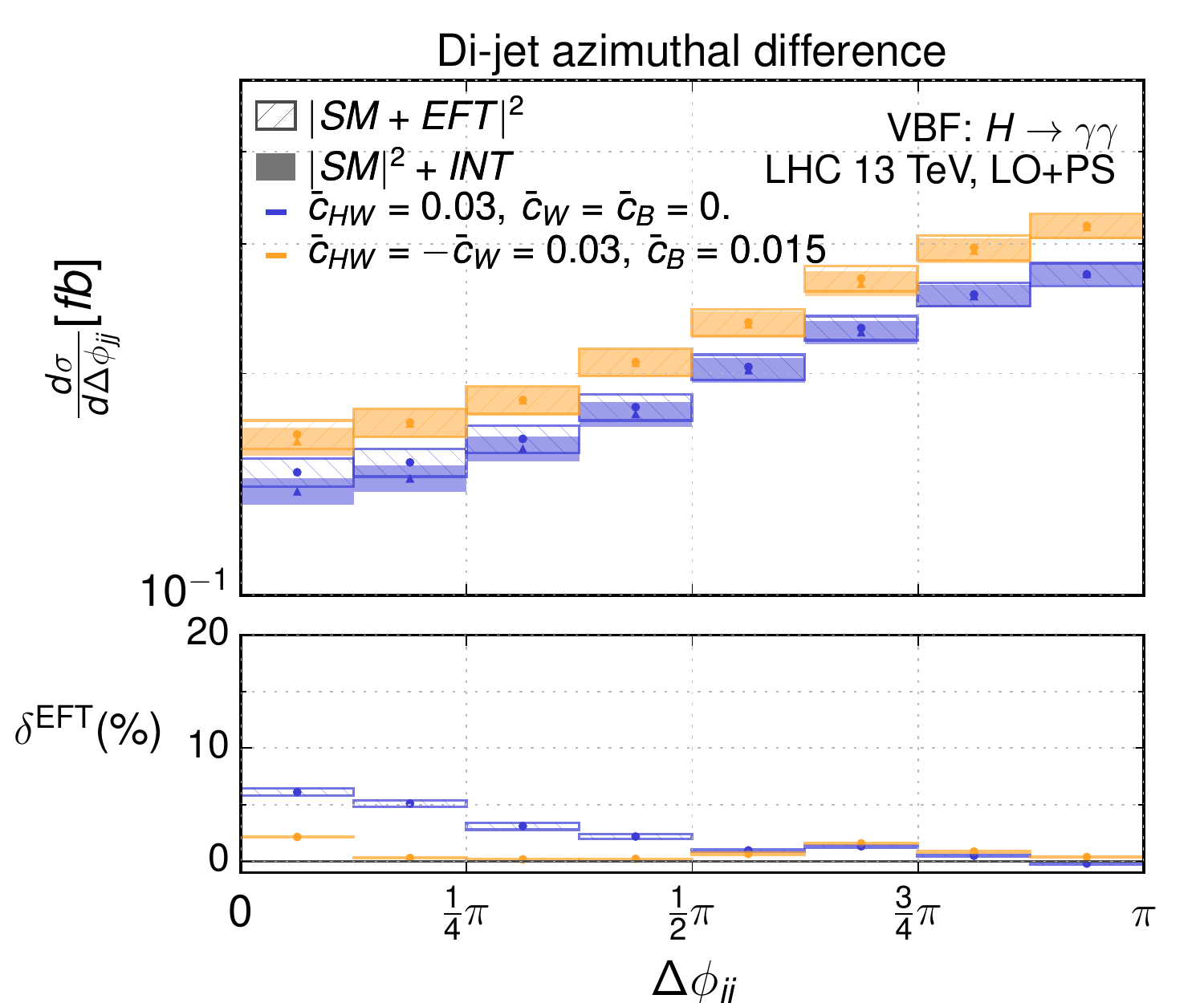}
  \caption{Same as in Fig.~\ref{fig:WH_INT}, but for VBF Higgs production.
 \label{fig:VBF} \label{fig:VBF_INT}}
\end{figure*}

We also repeat the simple EFT validity analysis performed for the $VH$ case and assess the impact of the EFT squared terms at LO, as shown in Fig.~\ref{fig:VBF_INT}. As suggested by the depletion effect of the EFT operators in the high energy bins of the differential distributions, the squared terms appear much more under control in this process compared to the $VH$ case. Within the ranges of our predictions the impact of the squared term is again most pronounced for the benchmark A, reaching at most \mbox{5--12\%} while for benchmark B their effect is much smaller.

\section{Future LHC reach}\label{sec:reach}
Before concluding, we attempt to estimate the reach of the LHC Run II with
respect to the Wilson coefficients considered in our benchmark scenarios. Our
results so far suggest to make use of the high energy tails of differential 
distributions as handles for new physics. For concreteness, we focus on the 
associated production process
\begin{align*}
    p\, p \to H \, W^\pm \to b \, \bar b \, \ell^\pm + {\slashed E}_T
\end{align*}
which has already been searched for at both LHC Run~I~\cite{ATLAS-CONF-2013-079}
and II~\cite{ATLAS-CONF-2016-091}. In both analyses, a large number of signal 
and control regions are defined according to the lepton and additional jet 
multiplicities, as well as to the vector boson transverse momentum $p_T^V$. These are combined in a global fit to obtain the corresponding SM Higgs signal strength. In
this fitting procedure, several dominant components of the background, namely $t\bar{t}$
and $W$-boson production in association with heavy-flavour jets, are left free 
to float. We consider as signal regions the $p_T^V$ overflow bin in the single 
lepton channel for both the zero-jet and one-jet categories. The Run~II study, 
however makes use of multivariate methods in the event selection process that 
are difficult to reproduce -- a task that definitely lies beyond the scope of the simple estimate we are intending to derive. We therefore choose to consider only the cut-based signal selection procedure employed in the Run~I analysis 
and then project the results for various Run~II integrated luminosities.

\subsection{Signal prediction and background estimation}
In order to estimate the number of background events in the single lepton signal
regions of a possible cut-based, LHC Run~II analysis, we extrapolate the 
results of the corresponding Run~I analysis. 
We first consider the dominating $t\bar{t}$ contribution
which arises from semi-leptonic top-antitop decays and which makes
up 54 and 85\% of the total background in the 0-jet and 1-jet categories, 
respectively. As a crude estimate for
the corresponding 13~TeV yields, we compute a transfer factor $f^i_{\text{tr}}$
(with $i=0$, 1 for the 0-jet and 1-jet categories)
by generating large statistics of SM semi-leptonic $t\bar{t}$ events at
centre-of-mass energies of 8 and 13 TeV on which we apply the kinematic
selection of the Run~I analysis summarised in Appendix~\ref{app:selection}. The
transfer factor $f^i_{\text{tr}}$ is defined as the ratio of the two fiducial
cross sections and we deduce the Run~II analysis background contributions by
multiplying the 8~TeV SM expectation $\sigma^i_{\text{bkg}}$ inferred from the
Run~I background event counts $N_{\text{bkg}}^i$ assuming 25 fb$^{-1}$ of 8 TeV
data. These should not depend much on the actual composition of 7 and 8 TeV data
analysed, particularly in the high transverse momentum overflow bin which is
dominated by 8 TeV data. 

Table~\ref{tab:ttbar_bkg} summarises the information 
obtained and used in the subsequent analysis. Our theoretical predictions for 
the $t\bar{t}$ contributions to the 0- and 1-jet signal regions at 8 TeV, 
$\sigma_{8}^{\text{overf.}}$, lie within a factor 2 of the cross-sections 
inferred from the post-selection, fitted background decomposition presented in 
Table 5 of Ref.~\cite{ATLAS-CONF-2013-079}. Due to the multi-variate nature of the recent Run II analysis, its fitted background yields cannot be used to validate the results of our projection, which rather represents the scenario in which a cut-based analysis similar to the Run I counterpart were performed at 13 TeV.

\begin{table}
    \begin{tabular}{c|cccc}
        \hline
        Category(i)&$N^i_{\text{bkg}}$
                &$\sigma_{8}^{\text{overf.}}$ 
                &$\sigma_{13}^{\text{overf.}}$ 
                &$f^i_{\text{tr}}$\tabularnewline
        \hline
        0-jet& 74  & 0.94  fb & 3.92 fb & 4.17  \tabularnewline
        1-jet& 143 & 3.20 fb  & 8.05 fb & 2.52  \tabularnewline
        \hline
    \end{tabular}
    \caption{
    Information necessary to estimate the 13~TeV projections of the
    fitted background yields, $N_{\text{bkg}}$, in the $p_T^V > 200$ GeV 
    overflow bins quoted in the analysis of Ref.~\cite{ATLAS-CONF-2013-079}. 
    We show the $t\bar{t}$ fiducial cross sections after the
    kinematic selection of Appendix~\ref{app:selection} for 8 and 13 TeV
    collisions and assuming a
    29.2\% semi-leptonic branching fraction, along with the 
    derived transfer factor $f^i_{\text{tr}}$.
    \label{tab:ttbar_bkg}}
\end{table}

The signal predictions have been generated using the previously described UFO
implementation, and both the $t\bar{t}$ and $WH$ contributions have been
simulated at the NLO accuracy in QCD as described in previous sections, the
fixed-order results being matched with {\sc {Pythia~8}} for both handling the
top decays and the parton showering. A grid of points spanning the allowed
region of the $(\bar{c}_{\sss HW},\,\bar{c}_{\sss W})$ parameter space, 
including the SM prediction, was simulated and the generated
events were passed through the same kinematic selection of
Appendix~\ref{app:selection}. Following the previous discussion on the existing
constraints, we assume the simplification $\bar{c}_{\sss W} =
-\bar{c}_{\sss B}/2$. Such a relation would not be retained in a complete global
fit including, e.g., LEP data. However, it is instructive to follow this
simplified path as it assesses the sensitivity of the LHC to the direction in
the $(\bar{c}_{\sss W},\,\bar{c}_{\sss B})$ plane that is orthogonal, and thus 
complementary, to the one tightly constrained by precision measurements at the 
$Z$-pole. We have derived least-squares-fitted quadratic
polynomial forms for the zero and one-jet overflow bin cross sections in the
two-dimensional parameter plane
$\sigma^i_{WH}(\bar{c}_{\sss HW},\bar{c}_{\sss W})$. Our results, moreover, 
embed a $b$-tagging efficiency of 70\%, and more information
(in particular on the explicit coefficients of the fits) is given in
Appendix~\ref{app:selection}.

\subsection{Results}
\begin{figure*}
\center
\includegraphics[width=0.325\textwidth]{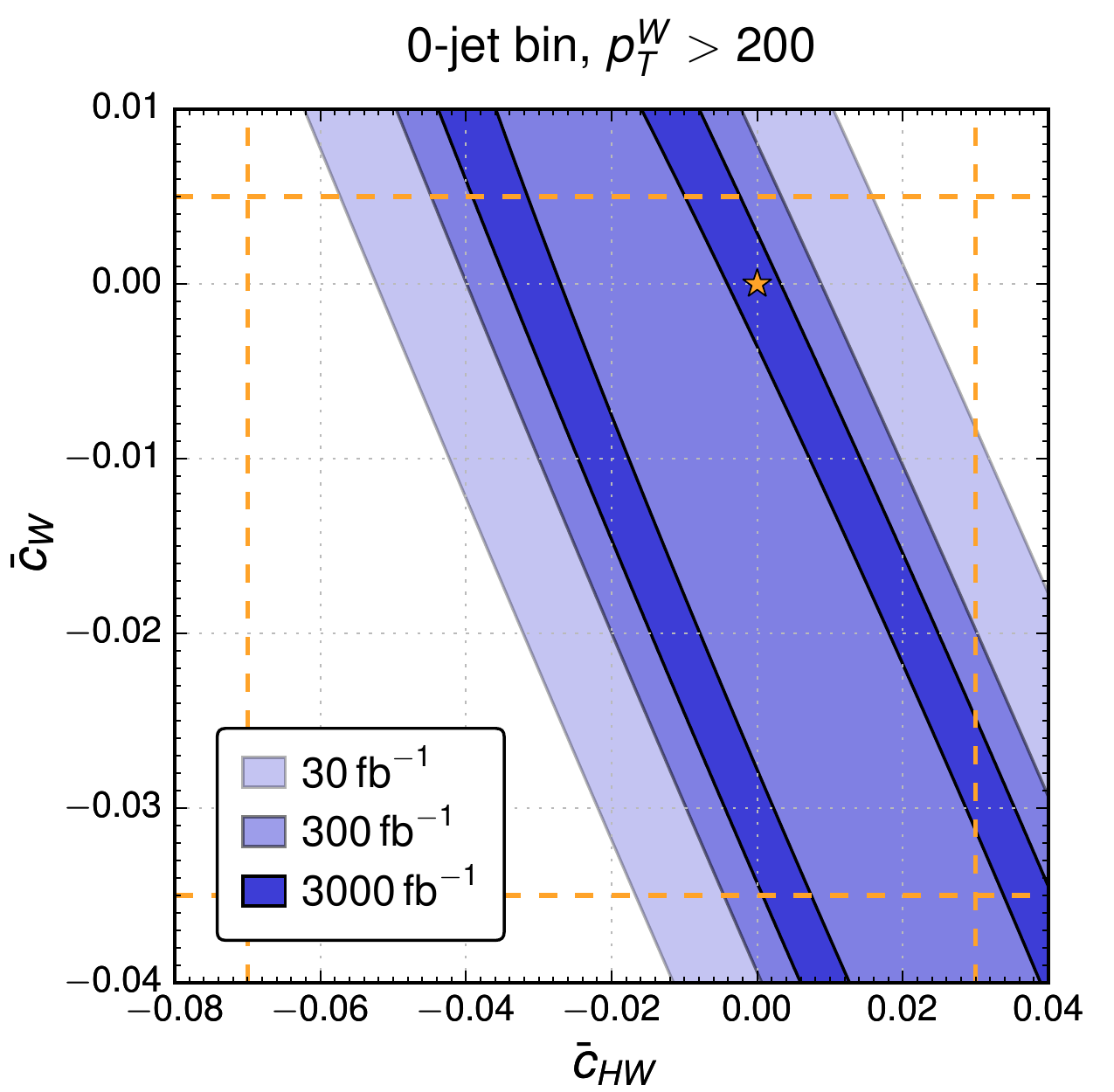}
\includegraphics[width=0.325\textwidth]{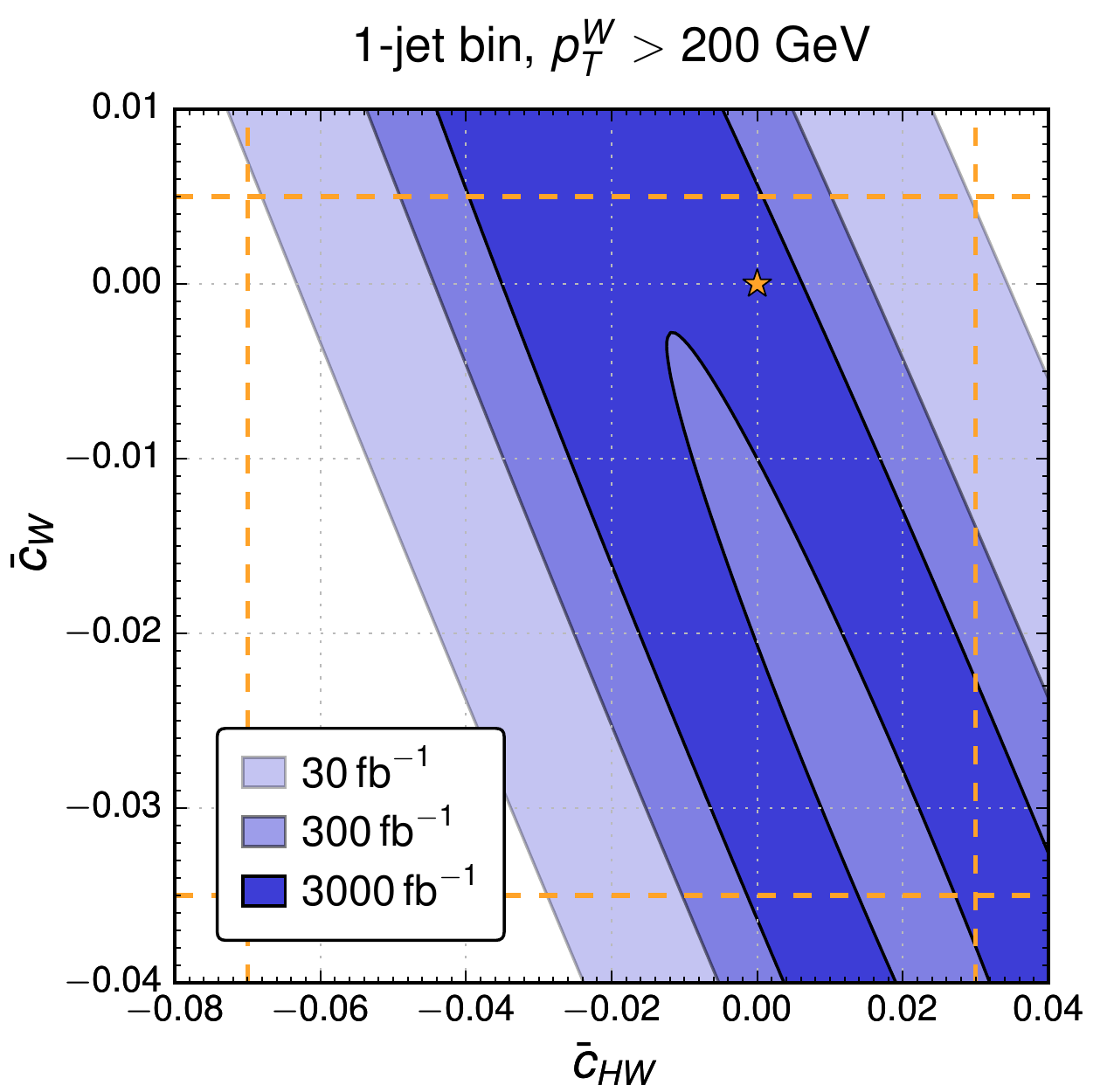}
\includegraphics[width=0.325\textwidth]{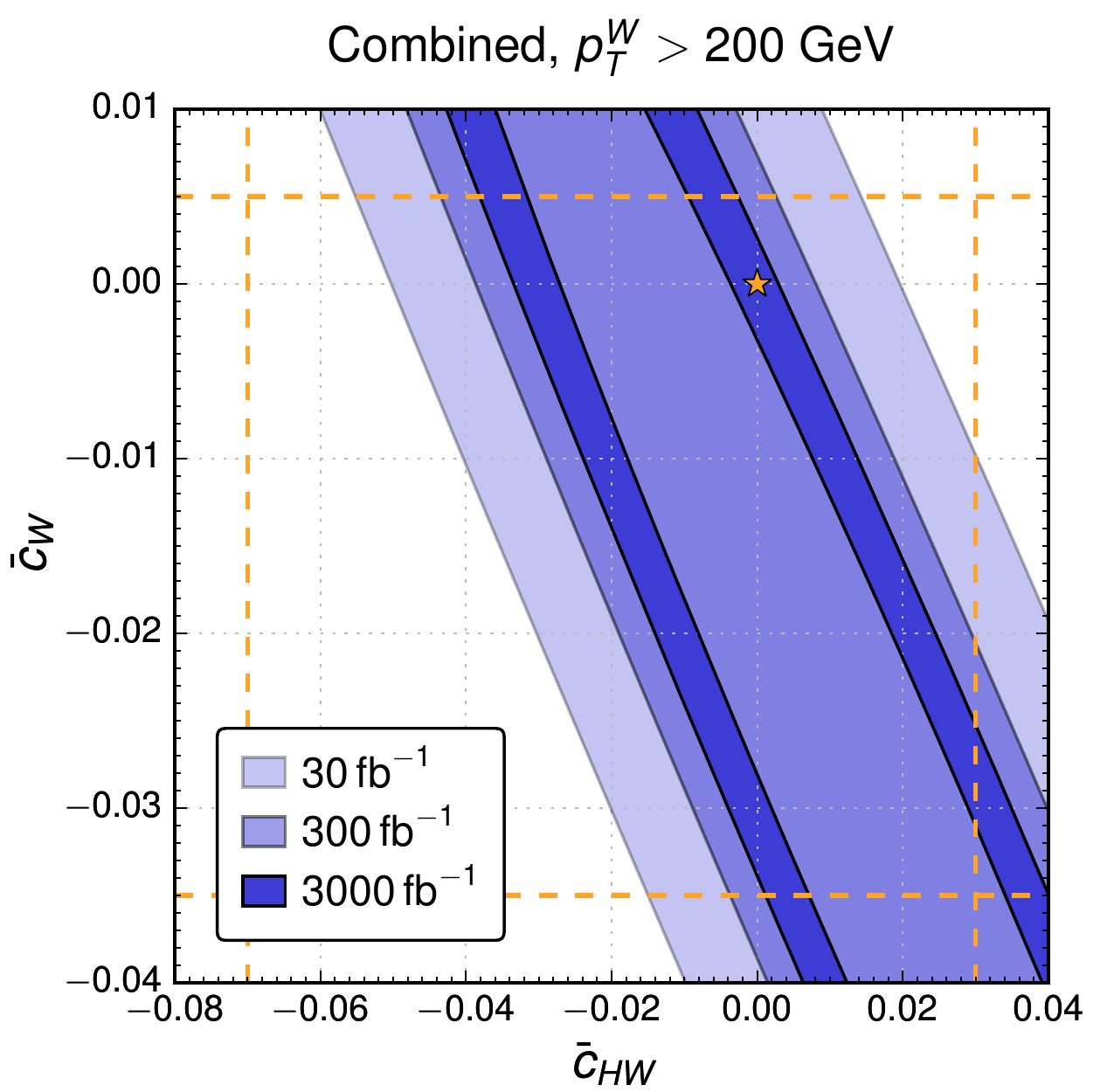}
  \caption{ 95\% confidence intervals in the
  $(\bar{c}_{\sss HW},\,\bar{c}_{\sss W})$ plane depicting the projected reach
  at the LHC Run~II extracted from data in the $p_T^W$ overflow bin of the
  corresponding Run~I analysis performed in Ref.~\cite{ATLAS-CONF-2013-079}. We
  consider three different choices for the integrated luminosities, and the
  dashed lines indicate the previously obtained marginalised limits on the Wilson coefficients from the global fit of Refs.~\cite{Ellis:2014jta,Ellis:2014dva}. \label{fig:reach}}
\end{figure*}

Our results have been derived from the fitted functional forms for the signal
cross sections in combination with the projected background
yields. We have performed a $\Delta\chi^2$ analysis to extract 95\% confidence
intervals assuming $\mathcal{L}=$30, 300 and 3000 fb$^{-1}$ of integrated
luminosities of 13~TeV proton-proton collisions. Denoting by $B^i$ and $S^i$ the
event counts in the signal region in the background-only and
signal-plus-background hypotheses, respectively, we have
\be\bsp
    B^i=&\mathcal{L}(f^i_{\text{tr}}\sigma^{i}_{\text{bkg}}
                    +\sigma^i_{WH}(0,0));
    \\
    S^i=&\mathcal{L}(f^i_{\text{tr}}\sigma^{i}_{\text{bkg}}
                    +\sigma^i_{WH}(\bar{c}_{\sss HW},\bar{c}_{\sss W}));
    \\
    \Delta\chi^2 \simeq &\sum_i
     \frac{(B^i-S^i)^2}
          {B^i},
    \\
    =&\sum_i
    \frac{\mathcal{L}\left[\sigma^i_{WH}(0,0)-
    \sigma^i_{WH}(\bar{c}_{\sss HW},\bar{c}_{\sss W})\right]^2}
         {f^i_{\text{tr}}\sigma^{i}_{\text{bkg}}
                    +\sigma^i_{WH}(0,0)}.
\esp\ee
The 95\% confidence intervals are obtained at the boundary of
$\Delta\chi^2=5.99$ which equates to the corresponding $p$-value for a $\chi^2$
distribution with two degrees of freedom. Fig.~\ref{fig:reach} depicts these
confidence intervals for the zero- and one-jet bin separately as well as their
combination, for the three integrated luminosity points. For comparison, the
marginalised single parameter exclusion regions established in Table~\ref{tab:LHCoperators} and the benchmark discussion of Sect.~\ref{ssec:bench} are indicated.

This simplified projection shows that this type of analysis is likely to
substantially improve the existing limits on these Wilson coefficients in
combination with existing data. Since the one-jet category suffers both from a
larger background and smaller signal contribution, its relative impact on the
overall reach is small. The blind direction associated with this measurement
lies very close to the $\bar{c}_{\sss HW}=-\bar{c}_{\sss W}$ line, corresponding
to the benchmark choice B of the earlier sections. This is consistent with the
very mild expected impact of this particular new physics scenario in the high 
energy tails of the differential distributions for the $p p \to W^+ \, H$ 
process (see Sect.~\ref{sec:WH}). Nevertheless, our results suggest that in 
the general case, taking the full integrated luminosity of LHC Run II will 
individually allow to constrain Wilson coefficients with a precision of a few 
per-mille and the results presented in Sec.~\ref{sec:VBF} indicate that 
combining VBF and WH studies will break this degeneracy.

\section{Conclusions}\label{sec:conclusion}

We have presented {\sc FeynRules} and UFO implementations of dimension-six SMEFT
operators affecting electroweak Higgs-boson production, which can be
used for NLO(QCD)+PS accurate Monte Carlo event generation within the
{\sc MG5\_aMC} framework. We have considered five SILH basis operators and have
accounted for all field redefinitions that are necessary to canonically
normalise the theory. Moreover, the ensuing modifications of both the gauge
couplings and the relationships between the electroweak inputs and the derived
parameters have also been included. We have showcased the strength of
our approach by simulating both associated VH and VBF Higgs-boson production at
the 13~TeV LHC, selecting a pair of benchmark scenarios informed both by recent
limits from global fits to the LEP and LHC Run~I data and by theoretical
motivations originating from integrating out certain popular ultraviolet
realisations.

We have found that EFT predictions and deviations from the
SM are stable under higher-order corrections. Overall, we have also observed a
significant reduction of the theoretical errors, which would have an impact on
the future measurements aiming to unravel dimension-six operator contributions.

Furthermore, as a test for the validity of the EFT approach, we have proposed to
compare distributions that either include the full matrix element (embedding all
SM and new physics contributions) or account solely for the interference of the
SM component with the new physics component.
For our benchmark choices that saturate current experimental limits, the differences were observed to be large in the kinematic extremes of some of our distributions, particularly for $VH$ production. This points to the possibility that the EFT description is breaking down in these regions of the parameter space and that the most precise measurements undertaken at the LHC Run II may be required to probe the EFT (while staying in its region of validity).

When comparing results for the VH and VBF channels, we have found that both
Higgs-boson production modes are sensitive to new physics, but the VH one seems
to have a better handle on $g^{(1)}$-type ($V_{\mu\nu}V^{\mu\nu}h$) structure, 
since several key distributions display deviations that may be more easily
distinguished from the background. Although the QCD $K$-factors have been
observed not to depend on the EFT parameters, the reduced theoretical
uncertainties are crucial for disentangling a non-vanishing
contribution of the $g^{(1)}$-type structure to the predictions from
the SM. This has been singled out in our study of the benchmark point B.
Moreover, our results exhibit an interesting complementarity of the two Higgs
production channels, since the interference pattern between the SM and the SMEFT 
contributions is quite different and benchmark-dependent.

In order to estimate the reach that might be possible at LHC Run II, we have
performed a simplified analysis projecting the Run I SM background expectations 
in a search for $WH$ associated production and combining this information with
LHC Run~II signal predictions obtained using our implementation. Using the 
overflow bin of the reconstructed $W$-boson transverse momentum distribution in 
the single lepton channel as a probe for EFT effects suggests that the LHC 
Run~II will significantly improve the current limits obtained from global fits.
Clearly both the $VH$ and VBF processes deserve further investigation including
detector effects and an analysis strategy to reject the SM backgrounds. In this 
case, the new physics contributions to the SM background processes should also 
correctly be accounted for, since effective operators affecting electroweak 
Higgs-boson production also impact the normalisation of the triple gauge-boson 
interactions both directly and indirectly via the aforementioned field 
redefinitions~\cite{Corbett:2013pja,Ellis:2014jta,Gorbahn:2015gxa,Mimasu:2015nqa,Butter:2016cvz}. 

Finally, our work has demonstrated a proof-of-concept for automated NLO+PS
simulations in the SMEFT framework. To this aim, we have limited ourselves to a
small set of dimension-six operators and a pair of benchmark points. This is
characterised as a first step towards a complete operator basis implementation,
in which the renormalisation group running of the Wilson coefficients~\cite{%
Elias-Miro:2013gya,Elias-Miro:2013mua,Jenkins:2013zja,Jenkins:2013wua,%
Alonso:2013hga,Maltoni:2016yxb} could also be supplemented in the future.

\section*{Acknowledgements}
We would like to express a special thanks to Fabio Maltoni for useful
discussions. We moreover acknowledge the organisers of the 2015 `Les
Houches--Physics at TeV colliders' workshop and the Mainz Institute for
Theoretical Physics for their hospitality and support during the
completion of this work. CD~is a Durham International
Junior Research Fellow. BF~and KMa~have been supported by the
Theory-LHC-France initiative of the CNRS (INP/IN2P3) and KMi~and VS~by
the Science and Technology Facilities Council (Grant number
ST/J000477/1).

\appendix
\begin{figure*}[h!]
\center
\includegraphics[width=0.325\textwidth]{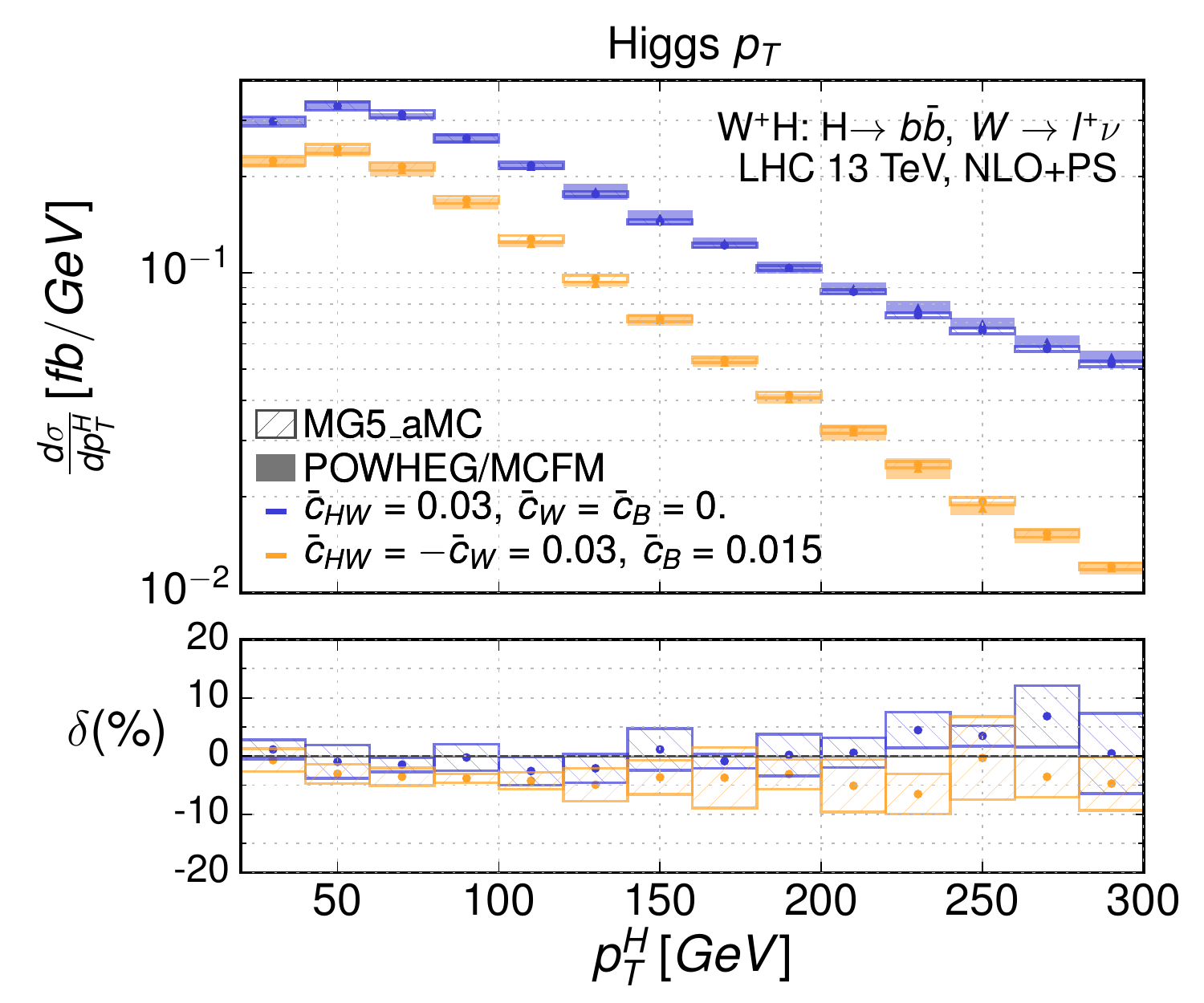}
\includegraphics[width=0.325\textwidth]{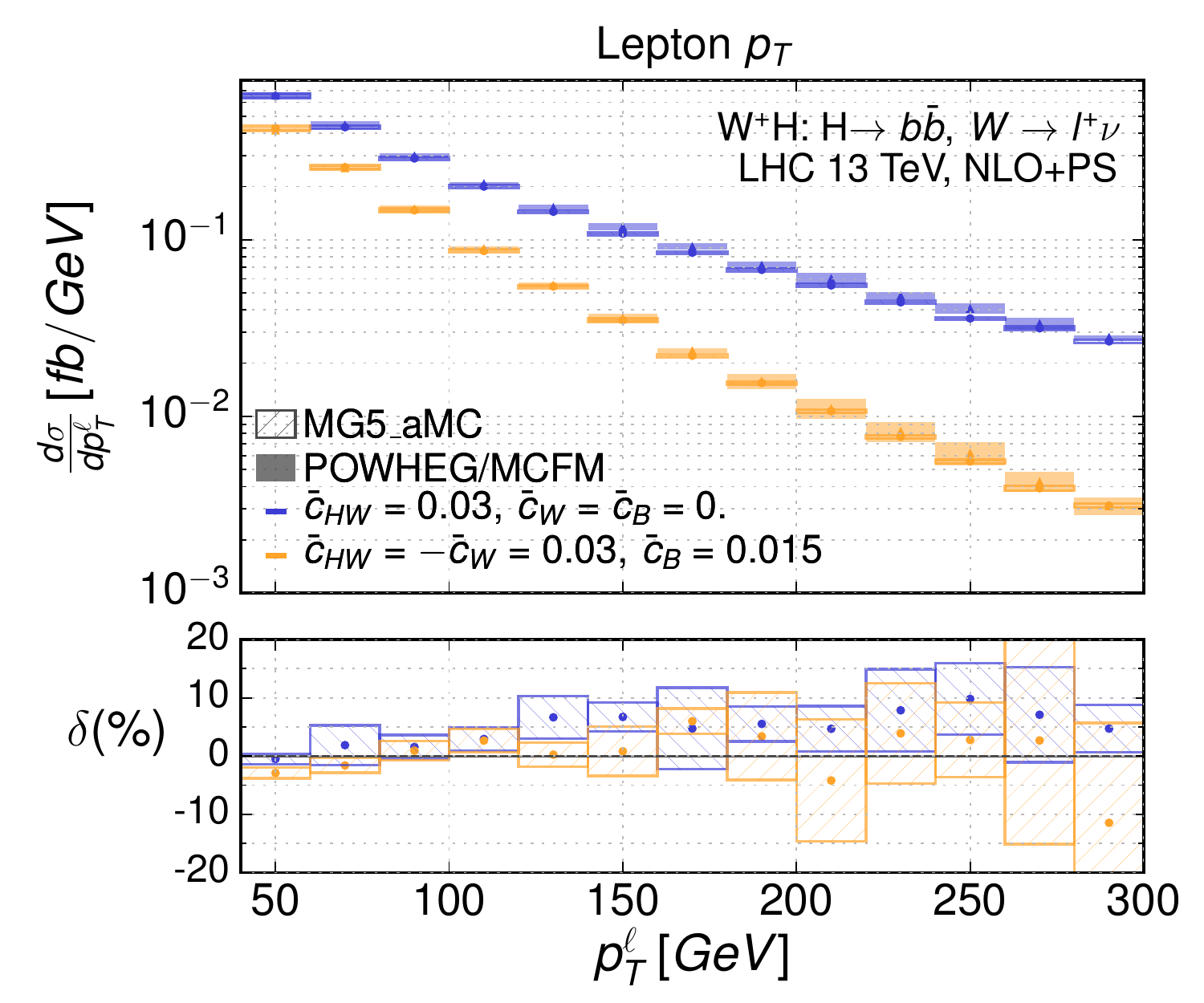}
\includegraphics[width=0.325\textwidth]{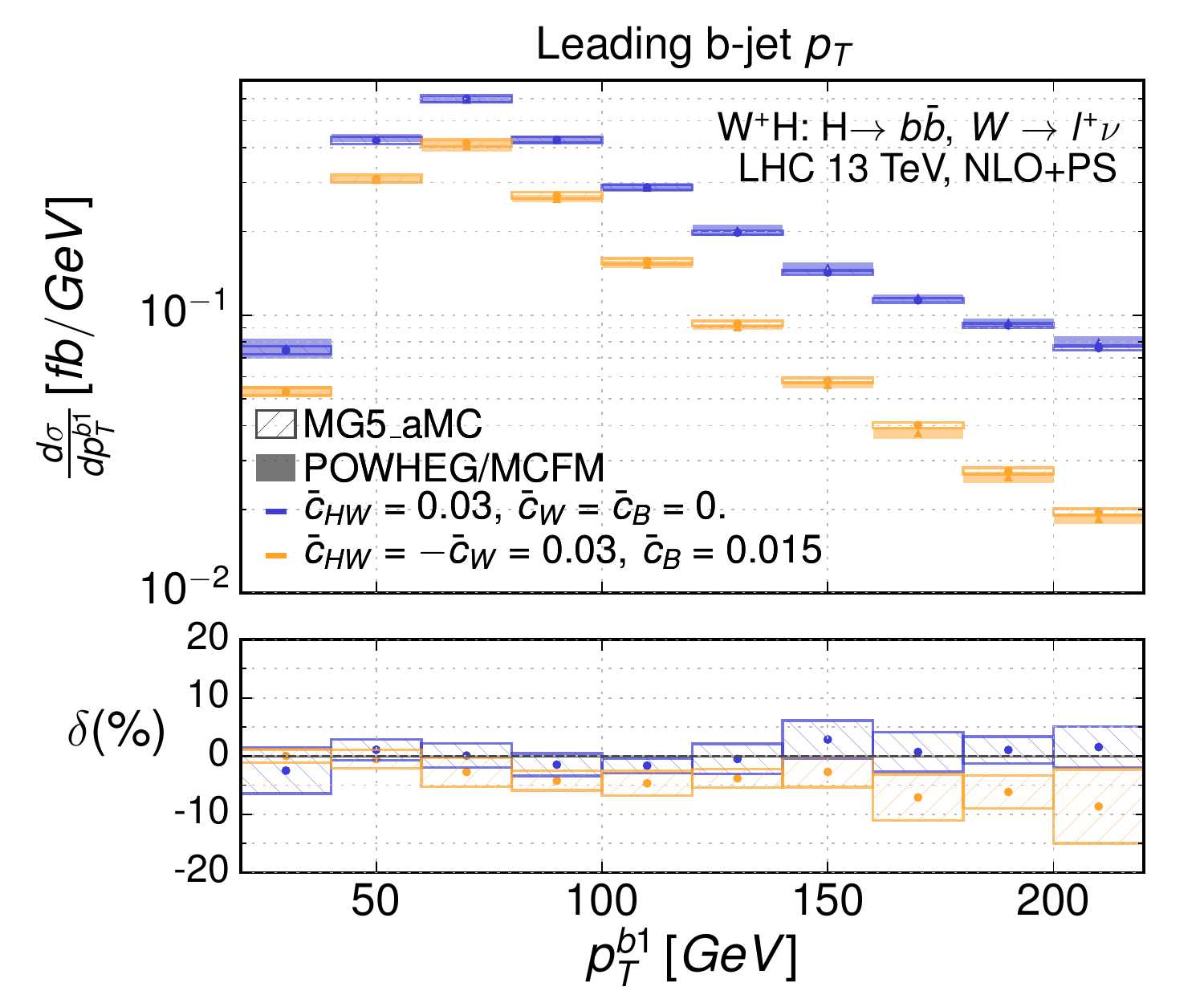}
  \caption{$W^+H$ differential distributions at NLO+PS in the {\sc MG5\_aMC} and
  MCFM/{\sc Powheg} frameworks.\label{fig:WH_CMP}}
\end{figure*}

\section{Simulation in {\sc MadGraph5\_aMC@NLO}\label{sec:implementation}}

\subsection{Technical details}

Our HEL@NLO UFO model can be downloaded from the {\sc FeynRules} model
database~\cite{FR-HEL:Online}.
It can be used for generating events at the NLO accuracy in QCD using software
such as {\sc MG5\_aMC} via to the automated procedure detailed in
Sect.~\ref{sec:sim}. 
Event generation for $W^+H$ production is achieved by typing in the {\sc MG5\_aMC}
interpreter
\begin{verbatim}
   import model HELatNLO
   generate p p > h ve e+ [QCD]
   output 
   launch
\end{verbatim}
Since the usual decay syntax of {\sc MG5\_aMC} is not
available for NLO event generation, we directly request the presence of the
$W$-boson decay products in the final state. An alternative way would require one to
simulate
the production of a Higgs boson in association with an on-shell $W$-boson that
is subsequently decayed within the {\sc MadSpin} infrastructure~\cite{%
Artoisenet:2012st,Alwall:2014bza} before invoking the parton showering.
On the other hand, VBF Higgs-boson production is achieved by typing in the
generation command
\begin{verbatim}
   generate p p > h j j $$ w+ w- z a QCD=0 [QCD] 
\end{verbatim}
The removal of the $s$-channel gauge boson contributions (via the {\tt \$\$}
syntax) avoids the generation of VH topologies where
the gauge boson decays hadronically.

In the parameter card, users may set values for the five
Wilson coefficients defined in Eq.~\eqref{eqn:lhelc} as well as for
the cutoff scale $\Lambda$. The modifications to the electroweak parameters in
terms of the inputs in the $(G_\mathrm{F}, \mz, \alpha)$ scheme are taken into account,
as well as the shifts induced by the field redefinitions discussed in
Sect.~\ref{sec:theory}.

At the {\sc FeynRules} level, the final Lagrangian involving all of the
redefined fields and parameters is expanded up to $\mathcal{O}(1/\Lambda^2)$.
From the point at which this truncation occurs, all subsequent performed
calculations will necessarily induce some $\mathcal{O}(1/\Lambda^n)$ (with
\mbox{$n > 2$}) dependence from, e.g., higher powers in the electroweak
couplings. Furthermore, {\sc MG5\_aMC} constructs its matrix elements by
squaring helicity amplitudes, so that the squared EFT contribution terms that
are formally of $\mathcal{O}(1/\Lambda^4)$ are by default retained on
top of the leading interference terms with the SM of $\mathcal{O}(1/\Lambda^2)$. A positive definite cross
section is ensured at the price of including higher-order terms. It is
therefore important to remain in the regime where the EFT expansion can be
trusted in that higher-order contributions are 
sufficiently suppressed. Departure from this safe zone may be reflected by a
rapid growth of amplitudes with energy leading to extreme deviations in the
tails of the distributions, as well as by discrepancies between independent
Monte Carlo setups performing their truncation in different ways. Alternatively, this
issue could be investigated  thanks to some recent developments in {\sc MG5\_aMC} that
allow the user to specify an order for the \textit{squared} matrix element
calculation, like {\tt QED\^{}2} or {\tt QCD\^{}2}. However, this feature is currently only available for LO computations.
For including the EFT effects in the simulation, the coupling order parameter
{\tt NP} can hence be set either to {\tt NP=1} to retain the full amplitude
squared or to {\tt NP\^{}2<=1} to throw away the aforementioned
EFT squared terms, e.g.
\begin{verbatim}
   generate p p > h ve e+ NP^2<=1
\end{verbatim}

Finally, due to the details of the implementation, we advise users seeking to
recover the SM limit to avoid setting the Wilson coefficients to zero. It is preferable to either set them to very small non-zero values or to use restrictions.
The cutoff scale parameter {\tt NPl} may also be alternatively fixed to a
very large value.


\subsection{Comparison with the MCFM/POWHEG-BOX implementation}

We have verified that our results in the $WH$ channel are compatible with those
stemming from an alternative existing implementation, which is based on
MCFM~\cite{Campbell:1999ah,Campbell:2011bn,Campbell:2015qma} and the
{\sc Powheg-Box} framework~\cite{Frixione:2007vw,Alioli:2010xd} and that has
been introduced in Ref.~\cite{Mimasu:2015nqa}. We have scrutinised several
differential distributions for both our benchmark points in the {\sc Powheg}
and {\sc MG5\_aMC} frameworks, as presented in Fig.~\ref{fig:WH_CMP}. A good
consistency has been found up to statistical uncertainties and despite the
difference in the dynamical scale choice which is taken as $H_T/2$ in
{\sc MG5\_aMC} and the invariant mass of the Higgs-vector boson system in the
{\sc Powheg-Box} implementation.

\section{Fit to kinematic selection of ATLAS-CONF-2013-079\label{app:selection}}
We summarise here the kinematic selection through which the $t\bar{t}$ and signal $WH$ events have been passed in order to determine the background transfer factor and signal efficiencies of the analysis performed in Ref.~\cite{ATLAS-CONF-2013-079}. The signal region used is the $p_T^W > 200$ GeV overflow bin in the 0 and 1-jet categories of the single-lepton channel. The kinematic selection for this channel applied to our event samples after parton shower is as follows:
\begin{itemize}
    \renewcommand{\labelitemi}{$\bullet$}
    \item We require the final state to contain exactly one lepton with $|\eta|<2.47$ and $E_T> 25$ GeV.
    \item Jets reconstructed by means of the anti-$k_T$ jet algorithm with a radius parameter $R=0.4$, and we discard the jet candidates for which the conditions
      $|\eta|<4.5$ and $p_T> 20$ GeV are not satisfied.
    \item We require exactly two $b$-jets with a pseudorapidity $|\eta|<2.5$,
     the hardest one being further imposed tho have a transverse momentum
    $p_T>$ 45 GeV.
\end{itemize}
Not more than one additional jet in the $|\eta|< 2.5$ region is allowed and any event with a jet with $p_T>$ 30 GeV in the $|\eta|> 2.5$ region is also rejected. Events are then split into the zero- and one-jet categories based on the presence of an extra, non-forward jet softer than the two $b$-jets.
The vector boson transverse momentum $p_T^W$ is defined as the vector sum of the transverse momentum of the lepton and the missing transverse energy. Additionally, in the $p_T^W > 200$ overflow bin, a cut of 50 GeV on the missing energy is imposed as well as a cut on the distance between the two $b$-jets, $\Delta R_{bb}\equiv\sqrt{\Delta \eta_{bb}^2+\Delta\phi_{bb}^2} < 1.4$. A flat $b$-tagging efficiency of 70\% is in addition assumed.

34 signal samples have been simulated with the Wilson coefficients being
taken in the ranges $-0.02 < \bar{c}_{\sss HW} < 0.03$ and $-0.03 <
\bar{c}_{\sss W} < 0.01$, the total rates being rescaled by the Higgs branching fraction to $b\bar{b}$ obtained with {\sc eHDECAY}. These events have then
been passed through the above selection in order to determine the functional forms for the zero- and one-jet signal cross sections in $p_T^W$ overflow bin, $\sigma^0$ and $\sigma^1$. A least-squares fit yields
\begin{align}
    \nonumber
    \sigma^0_{WH} =& 1.135(1 + 71\,\bar{c}_{\sss HW} + 79.5\,\bar{c}_{\sss W}
                + 228\,\bar{c}_{\sss HW}^2 \\
                &\qquad \,\,\,\,\,+ 253\,\bar{c}_{\sss W}^2 
                + 4711\,\bar{c}_{\sss HW}\,\bar{c}_{\sss HW})
                \text{ fb},\\\nonumber
    \sigma^1_{WH} =& 0.607(0.9 + 66.8\,\bar{c}_{\sss HW} + 74.6\,\bar{c}_{\sss W}
                + 232\,\bar{c}_{\sss HW}^2 \\
                &\qquad \quad \,\,\,\,+ 243\,\bar{c}_{\sss W}^2 
                + 4621\,\bar{c}_{\sss HW}\,\bar{c}_{\sss HW})\text{ fb}.
\end{align}
The fit coefficients are within 5--10\% of one another, which is to be expected given that our signal process receives contributions from an electroweak vertex and should not be directly sensitive to additional QCD radiation.

\bibliography{biblio}
\bibliographystyle{JHEP}

\end{document}